\documentclass[12pt]{article}
\begin{document}
\vspace*{2cm} \noindent {\Large\bf Spherical Model in a Random
Field.} \vspace{\baselineskip}
\newline {\bf A.E.\ Patrick\footnote[1]{Laboratory of Theoretical
Physics, Joint Institute for Nuclear Research, Dubna 141980,
Russia e-mail: patrick@theor.jinr.ru}}
\begin{list}{}{\setlength{\rightmargin }{0 mm}
\setlength{\leftmargin }{2.5cm}} \item \rule{124mm}{0.3mm}
%\item \rule{89mm}{0.3mm}
\linebreak {\footnotesize {\bf Abstract.} We investigate the
properties of the Gibbs states and thermodynamic observables of
the spherical model in a random field. We show that on the
low-temperature critical line the magnetization of the model is
not a self-averaging observable, but it self-averages
conditionally. We also show that an arbitrarily weak homogeneous
boundary field dominates over fluctuations of the random field
once the model transits into a ferromagnetic phase. As a result, a
homogeneous boundary field restores the conventional
self-averaging of thermodynamic observables, like the
magnetization and the susceptibility. We also investigate the
effective field created at the sites of the lattice by the random
field, and show that at the critical temperature of the spherical
model the effective field undergoes a transition into a phase with
long-range correlations $\sim r^{4-d}$. }
\newline \rule[1ex]{12.4cm}{0.3mm}
%\newline \rule[1ex]{89mm}{0.3mm}
\linebreak {\bf \sc key words:} {\footnotesize Critical
fluctuations; disordered spin systems; Gibbs states;
self-averaging.} \vspace{\baselineskip}
%\newline PACS.\ 75.10H
\end{list}
%\vspace{\baselineskip}

\section{Introduction.}
The spherical model \cite{bk} is a lattice model where a
(thermodynamic) random variable $x_j$ is attached to every site
$j$ of a subset $V_n$ of a $d$-dimensional square lattice $Z^d$.
This model is one of a handful of models where exact results can
be obtained in the presence of a random field $\{h_j, j\in Z^d\}$.
Thermodynamic properties of such a disordered spherical model
outside the low-temperature critical line were studied by Pastur
in the paper \cite{p82}. The magnetization on the critical line
was also derived there in the limits $h_0\to\pm0$, where $h_0$ is
the expected value of the random field.

Some thermodynamic characteristics have discontinuities on the
critical line, and, depending on the boundary conditions and the
exact details of passing to the thermodynamic limit ($V_n\uparrow
Z^d$), those characteristics can have different limiting values.
Their values in the limits $h_0\to\pm0$ are, in some sense,
extreme points of the sets of all possible limiting values. For
some models those sets contain simply all linear combinations of
the extreme values. For disordered models, like the spherical
model in a random field, that is not necessarily the case. The aim
of this paper is to study thermodynamic properties of the
spherical model directly on the low-temperature critical line.

Many models in statistical mechanics are complicated enough to
force us to restrict the investigation to finding only certain
thermodynamic averages. For instance, sometimes investigation of
magnetization is reduced to calculation of the averages
\[
\langle m_N\rangle=\frac{1}{N}\sum_{j\in V}\langle x_j\rangle,
\]
where $\langle\cdot\rangle$ denotes the average over the Gibbs
distribution. However, as a rule, for a satisfactory understanding
of properties of a particular model (especially on a critical
line) one has to know distributions of various macroscopic (and,
ideally, microscopic) quantities. For that reason in the present
paper we study the limiting {\em Gibbs states} and the {\em
distributions} of thermodynamic observables.

One of the properties particular to disordered systems in
statistical mechanics is the {\em self-averaging\/} of
thermodynamic observables, introduced by Pastur and Figotin in the
paper \cite{pf78}. There they also proved a general theorem
concerning the self-averaging of thermodynamic observables for a
wide class of models. By observables they meant quantities already
averaged over the Gibbs distribution. For disordered systems
involving a (realization of a) random field $\{h_j,j\in Z^d\}$ the
self-averaging is defined as follows. \vspace{\abovedisplayskip}

{\sf Definition 0} (see \cite{pf78}).  A thermodynamic observable
$\langle Q_N\rangle$ is self-averaging, if
\begin{equation}
\lim_{N\to\infty}\langle Q_N\rangle=Q \label{sav0}
\end{equation}
exists and is the same for almost all realizations of the random
field, where
%. In the above formula, the brackets $\langle\cdot\rangle$ denote
%thermodynamic average (w.r.t.\ the Gibbs distribution), and
$N$ is the size of the system. \vspace{\abovedisplayskip}

The name self-averaging indicates that one does not have to
average the thermodynamic observable $Q_N$ over the distribution
of the random field. Indeed the limiting distribution is
concentrated at the average value, since Eq.\ (\ref{sav0})
trivially implies
\[
\lim_{N\to\infty}\langle Q_N\rangle=\mbox{\boldmath $E$}Q,
\]
where $\mbox{\boldmath $E$}(\cdot)$ denotes the average over the
distribution of the random field. As a rule self-averaging
observables are uniformly integrable, see \cite{b00,s98}, hence,
it is also true that
\[
\lim_{N\to\infty}\mbox{\boldmath $E$}\langle Q_N\rangle=Q.
\]

From probabilistic point of view there are no fundamental
differences between the thermodynamic randomness (described by the
Gibbs distribution) and the randomness of the field $\{h_j,j\in
Z^d\}$. Therefore it seems natural to get rid of the thermodynamic
averages in the definition of self-averaging for observables like
the magnetization. \vspace{\abovedisplayskip}

{\sf Definition 1}.  A thermodynamic observable $Q_N$ is
self-averaging, if
\begin{equation}
\lim_{N\to\infty} Q_N=Q, \label{sav1}
\end{equation}
exists and is the same for almost all realizations of the random
field $\{h_j,j\in Z^d\}$, where the limit is understood in
probability w.r.t.\ the thermodynamic randomness.
\vspace{\abovedisplayskip}

%\noindent In this paper we always use Definition 1 (as opposed to
%Definition 0) as the definition of self-averaging.
There are thermodynamic observables which are not self-averaging
on critical lines/points, having continuous (non-thermodynamic)
distributions. For instance, it is widely known that the
susceptibility
\[
\chi_N=\frac{1}{N}\sum_{j,k=1}^N \langle x_jx_k\rangle-\langle
x_j\rangle \langle x_k\rangle
\]
is an observable of that kind. On the other hand, there are
observables which distributions concentrate at a few (two or more)
points. This fact motivated the authors of the paper \cite{apz92}
to introduce the notion of the conditional
self-averaging.\vspace{\abovedisplayskip}

{\sf Definition 2} (see \cite{apz92}). A thermodynamic observable
$Q_N$ is {\em conditionally self-averaging}, if
\begin{equation}
\lim_{N\to\infty}Q_N-\mbox{\boldmath
$E$}(Q_N|\xi_N)=0,\qquad\mbox{in probability,} \label{csav}
\end{equation}
where $\mbox{\boldmath $E$}(\cdot|\xi_N)$ are the conditional
averages w.r.t.\ a sequence of functions of the random field
$\{\mbox{\boldmath $h$}_j,j\in Z^d\}$ which obtain only a finite
number of values, $F$, the same for all $N$.
\vspace{\abovedisplayskip}

For an illustration of the notion of conditional self-averaging
one can look at the random-field Curie-Weiss model, see
\cite{apz92}. In this model a conditionally self-averaging
observable $Q_N$ is the magnetization
$\frac{1}{N}\sum_{j=1}^{N}s_j$, the sequence of functions $\xi_N$
is the sign of the total random field
\[
\xi_N={\rm sgn}\left( \sum_{j=1}^Nh_j\right),
\]
and  $\mbox{\boldmath $E$}(Q_N|\xi_N)\sim\xi_Nm^*$, where $m^*$ is
the spontaneous magnetization.

For a self-averaging observable $Q_N$ both thermodynamic
(described by the Gibbs distribution) and non-thermodynamic
(produced by the random field) fluctuations vanish as
$N\to\infty$. It seems useful to introduce exponents which
indicate how fast that happens. The exponent $\rho$, related to
non-thermodynamic fluctuations, is defined by
\begin{equation}
\left\langle Q_N-\mbox{\boldmath $E$}\langle
Q_N\rangle\right\rangle= N^{-\rho}r_N, \label{expr}
\end{equation}
as $N\to\infty$, where the sequence of random variables $r_N$
converges to a random variable with a proper, non-degenerate
distribution. The exponent $\tau$, indicating the magnitude of
thermodynamic fluctuations, is defined by
\begin{equation}
Q_N-\mbox{\boldmath $E$}\langle Q_N\rangle-N^{-\rho}r_N=
N^{-\tau}t_N, \label{expt}
\end{equation}
as $N\to\infty$, where, again, the sequence of random variables
$t_N$ converges to a random variable with a proper, non-degenerate
distribution. The definitions of exponents $\rho$ and $\tau$
generalize straightforwardly to the case of conditional
self-averaging.

As a rule thermodynamic systems outside critical lines/points are
collections of random variables $\{x_j\}_{j=1}^N$ with short-range
correlations. In this case one usually has self-averaging with the
exponents $\rho=\tau=\frac{1}{2}$. More precisely,
\[
m_N=\frac{1}{N}\sum_{j=1}^N x_j=x+N^{-\frac{1}{2}}r_N
+N^{-\frac{1}{2}}t_N.
\]
The exponents $\rho$ and $\tau$ are not fundamentally novel
quantities. For most commonly used thermodynamic observables $Q_N$
they are related in some way to the standard critical exponents.
The values of exponents for the magnetization of the spherical
model are calculated in this paper.

Somewhat different terminology was used in the papers
\cite{ah96,wd95}. There self-averaging with exponents
$\rho=\tau=\frac{1}{2}$ is called strong self-averaging, while
self-averaging with exponents $\rho\in(\frac{1}{2},1)$ and
$\tau\in(\frac{1}{2},1)$ is called weak self-averaging.

Some general results on the behaviour of models under the
influence of  random field were obtained in the 70s and 80s by
application of the renormalization-group ideas to the
Ginzburg-Landau model, see \cite{aim76,im75}. In particular it was
noticed that the random-field fluctuations dominate over the
thermodynamic fluctuations as the critical point is approached.
This observation suggests that the random-field fluctuations also
dominate on the low-temperature critical line, and hence one
should have $\tau>\rho$ there. This is exactly what happens with
the fluctuations of the magnetization of the spherical model, and
we will see in Section 6 that in this case
$\rho=\frac{1}{2}-\frac{2}{d}$ and $\tau=\frac{1}{2}-\frac{1}{d}$.

The rest of the paper is organized as follows. Section 2 contains
the exact definition of the spherical model, the random field and
the boundary conditions. It also contains some well known
technical results for the use in the later sections. Section 3
summarizes the main results of the paper. In Section 4 we
calculate the free energy of the spherical model as an
illustration of the application of saddle-point method in the
low-temperature region. In Section 5 we describe in details the
properties of the spherical model (the random field $\{x_j, j\in
Z^d\}$) in the infinite-volume limit. In Section 6 we provide an
analogous detailed description for the magnetization of the
spherical model. The results of Sections 5 and 6 in the absence of
the boundary field are re-derived in Section 7. The results of the
paper are discussed in Section 8.

\section{The model and useful facts.}
The spherical model describes a collection of random variables
$\{x_j,j\in Z^d\}$ placed at sites of an integer $d$-dimensional
lattice, $Z^d$. Every site $j\in Z^d$ is specified by its $d$
integer coordinates $(j_1,j_2,...,j_d)$.

To define the distribution of random variables at all sites of the
lattice, we first specify the joint distribution for the random
variables in a finite rectangle
\[
V_n=\{j\in Z^d:1\leq j_\nu\leq n,\nu=1,2,...,d\/\}
\]
containing $N\equiv n^d$ sites, and then pass to the limit
$n\to\infty$. To avoid unnecessary complications we impose
periodic boundary conditions in dimensions $2,3,\ldots,d$. Thus
the boundary of the rectangle $V_n$ is the set
\[
B_n=\{j\in V_n: j_1=1,n\}.
\]

\underline{\bf The Hamiltonian.} \vspace{\abovedisplayskip}

\noindent The random variables located in the rectangle $V_n$
interact with the boundary field, the external random field, and
each other via the Hamiltonian
\[
H_n=-J\sum_{j,k\in V_n}T_{jk}x_j x_k -\sum_{j\in V_n}h_j
x_j-b\sum_{j\in B_n} x_j,
\]
where $J>0$, $T_{jk}$ are the elements of the nearest-neighbour
interaction matrix, $\{h_j,\,j\in Z^d\}$ is a fixed realization of
the external random field, and $b$ is the boundary field.
\vspace{\abovedisplayskip}

\underline{\bf The interaction matrix.} \vspace{\abovedisplayskip}

\noindent The elements of the interaction matrix $\widehat{T}$ are
given by
\[
T_{jk}=\sum_{\nu=1}^d
J^{(\nu)}(j_\nu,k_\nu)\prod_{l\in\{1,2,\ldots,d\}\setminus\nu}\delta(j_l,k_l),
\]
where
\[
\delta(j_l,k_l)=\left\{
\begin{array}{cl}1,&\mbox{ if }j_l=k_l,\\
0,&\mbox{ if }j_l\neq k_l,
\end{array}
\right.
\]
is the Kronecker delta.

The coefficients $J^{(1)}(j_1,k_1)$ are the elements of the
$n\times n$ tri-diagonal matrix
\[
\widehat{J}^{(1)}=\left(
\begin{array}{ccccccc}0&\frac{1}{2}&&&&&\\
\frac{1}{2}&0&\frac{1}{2}&&&\mbox{\LARGE 0}&\\
&\frac{1}{2}&0&\ddots&&&\\
&&\ddots&\ddots&\ddots&&\\
&&&\ddots&0&\frac{1}{2}&\\
&\mbox{\LARGE 0}&&&\frac{1}{2}&0&\frac{1}{2}\\
&&&&&\frac{1}{2}&0
\end{array}
\right).
\]
The coefficients $J^{(\nu)}(j_\nu,k_\nu)$, for $\nu=2,3,\ldots,d$,
are the elements of the matrices $\widehat{J}^{(\nu)}$ which have
extra $\frac{1}{2}$ at the upper right and lower left corners (due
to the periodic boundary conditions)
\[
\widehat{J}^{(\nu)}=\left(
\begin{array}{ccccccc}0&\frac{1}{2}&&&&&\frac{1}{2}\\
\frac{1}{2}&0&\frac{1}{2}&&&\mbox{\LARGE 0}&\\
&\frac{1}{2}&0&\ddots&&&\\
&&\ddots&\ddots&\ddots&&\\
&&&\ddots&0&\frac{1}{2}&\\
&\mbox{\LARGE 0}&&&\frac{1}{2}&0&\frac{1}{2}\\
\frac{1}{2}&&&&&\frac{1}{2}&0
\end{array}
\right).
\]
The eigenvalues of the matrix $\widehat{J}^{(1)}$ are given by
\[
\Lambda_l=\cos\frac{\pi l}{n+1},\quad l=1,2,\ldots,n.
\]
The corresponding orthonormal (that is, orthogonal and normalised)
eigenvectors are given by
\[
\mbox{\boldmath $v$}^{(l)}
=\left\{v_m^{(l)}=\sqrt{\frac{2}{n+1}}\sin\frac{\pi l
m}{n+1}\right\}_{m=1}^n,\quad l=1,2,\ldots,n.
\]
The eigenvalues and orthonormal eigenvectors of the matrices
$\widehat{J}^{(\nu)}$, for $\nu=2,3,\ldots,d$, are given by
\[
\lambda_l=\cos\frac{2\pi (l-1)}{n},\quad l=1,2,\ldots,n,
\]
and
\[
\mbox{\boldmath $u$}^{(l)}
=\left\{u_m^{(l)}=\sqrt{\frac{2}{n}}\cos\left[\frac{2\pi (l-1)
(m-1)}{n}-\frac{\pi}{4}\right]\right\}_{m=1}^n,\quad
l=1,2,\ldots,n.
\]
Finally, the eigenvalues of the interaction matrix $\widehat{T}$
are the sums of the eigenvalues of the matrices
$\widehat{J}^{(\nu)}$
\[
\lambda_k=\Lambda_{k_1}+\sum_{\nu=2}^d\lambda_{k_\nu},\quad
k\equiv(k_1,k_2,\ldots,k_d)\in V_n.
\]
The corresponding orthonormal eigenvectors are the products of the
eigenvectors of the matrices $\widehat{J}^{(\nu)}$
\begin{equation}
\mbox{\boldmath $w$}^{(k)}
=\left\{w_j^{(k)}=v_{j_1}^{(k_1)}\prod_{\nu=2}^d
u_{j_\nu}^{(k_\nu)}\right\}_{j\in V_n},\quad
k\equiv(k_1,k_2,\ldots,k_d)\in V_n.
\label{evs}
\end{equation}
%\vspace{\abovedisplayskip}

\underline{\bf The external random field.}
\vspace{\abovedisplayskip}

We assume that the coefficients $\{h_j,j\in Z^d\}$ are a fixed
realization of independent normal random variables
$\{\mbox{\boldmath $h$}_j,j\in Z^d\}$ with zero mean and variance
$h^2$. The assumptions of independence and normal distribution are
made to avoid unnecessary complications. The behavior of the model
is very different if the random variables $\{\mbox{\boldmath
$h$}_j,j\in Z^d\}$ have, say, Cauchy distribution, or, if the
random variables have strong negative correlations severely
suppressing fluctuations of sums like $\sum_{j\in
V_n}\mbox{\boldmath $h$}_j$. Nevertheless, we restrict our
attention to the technically convenient case of independent normal
random variables where the fluctuations are neither abnormally
large, nor abnormally small. \vspace{\abovedisplayskip}

\underline{\bf The Gibbs distribution.} \vspace{\abovedisplayskip}

The distribution of the thermodynamic random variables $\{x_j,j\in
V_n\}$ is specified by the usual Gibbs density
\[
p(\{x_j,j\in V_n\})=\frac{e^{-\beta H_n}}{\Theta_n},
\]
with respect to the spherical {\em ``a priori''} measure
\[
\mu_n(dx)=\delta\left(\sum_{j\in V_n}x_j^2-N\right)\prod_{j\in
V_n}dx_j.
\]
The normalization factor (partition function) $\Theta_n$ is given
by
\begin{equation}
\Theta_n=\int_{-\infty}^\infty\ldots\int_{-\infty}^\infty
e^{-\beta H_n}\mu_n(dx).
\label{pf}
\end{equation}
%\vspace{\abovedisplayskip}

\underline{\bf Useful estimates.} \vspace{\abovedisplayskip}

\noindent Equations (\ref{wf})--(\ref{gf}) below state well known
results which are used throughout the paper. A routine analysis of
the singularity at $\omega_1=\omega_2=\ldots=\omega_d=0$ shows
that the function
\begin{equation}
W_d^{(m)}(z)\equiv\int_{-\pi}^{\pi}\ldots \int_{-\pi}^{\pi}
\frac{1}{\left(z-\sum_{\nu=1}^d \cos
\omega_\nu\right)^{m}}\prod_{\nu=1}^d \frac{d\omega_\nu}{2\pi}<
\infty \label{wf}
\end{equation}
at $z=d$ if $d>2m$.

Let $\gamma\in[0,2)$, $\zeta>0$, and $z_n=\lambda_{\rm max}+\zeta
n^{-\gamma}$, then we have as $n\to\infty$
\begin{equation}
\frac{1}{N}{\sum_{k\in
V_n}}\frac{1}{\left(z_n-\lambda_k\right)^m}=W_d^{(m)}(z_n)-
\frac{1}{2n}\triangle W_d^{(m)}(z_n)
+\,o\left[\exp\left(-n^{1-\gamma/2}c(\zeta)\right)\right],
\label{stoi0}
\end{equation}
where
\[
\triangle W_d^{(m)}(z_n)\equiv
W_{d-1}^{(m)}(z_n-1)+W_{d-1}^{(m)}(z_n+1)- 2W_d^{(m)}(z_n),
\]
and $c(\zeta)$ is strictly positive and increasing for $\zeta>0$.

If $\gamma=2$, $\zeta\geq 0$, and $d>4$, then
\begin{equation}
\frac{1}{N}{\sum_{k\in V_n}}'\frac{1}{\lambda_{\rm max}+\zeta
n^{-2}-\lambda_k}=W_d^{(1)}(d)- \frac{1}{2n}\triangle W_d^{(1)}(d)
-\zeta W_d^{(2)}(d)\,n^{-2}+o(n^{-2}),
%\int_{-\pi}^{\pi}\ldots \int_{-\pi}^{\pi}
%\frac{1}{\left(\lambda_{\rm max}+\zeta n^{-\gamma}-\sum_{\nu=1}^d
%\cos \omega_\nu\right)^{m}}\prod_{\nu=1}^d
%\frac{d\omega_\nu}{2\pi},
\label{stoi}
\end{equation}
as $n\to\infty$, where the prime indicates that the summation does
not involve $k=(1,1,\ldots,1)$.

If $d>2m$, and $\zeta\geq 0$, then
\begin{equation}
\frac{1}{N}{\sum_{k\in V_n}}'\frac{1}{\left(\lambda_{\rm
max}+\zeta n^{-2}-\lambda_k\right)^m}=W_d^{(m)}(d)+o(1),
\label{stoim}
\end{equation}
as $n\to\infty$. Approximation of sums of the type (\ref{stoi0}),
(\ref{stoi}) by integrals was analysed in \cite{bf73,fp86}. For an
outline of a method particularly suited for the above sums see
\cite{p94}.

If $m>0$ and $d>2m$, then
\begin{equation}
\int_{-\pi}^{\pi}\ldots\int_{-\pi}^{\pi}
\frac{\exp\left(i\sum_{\nu=1}^d x_\nu\omega_\nu\right)}
{\left(d-\sum_{\nu=1}^d \cos \omega_\nu\right)^m}\prod_{\nu=1}^d
\frac{d\omega_\nu}{2\pi}\sim\frac{\Gamma(d/2-m)}{2^m\pi^{d/2}\Gamma(m)}
\left(\sum_{\nu=1}^d x_\nu^2\right)^{m-d/2}, \label{gf}
\end{equation}
as $\sum_{\nu=1}^d x_\nu^2\to\infty$. For a derivation of the
above asymptotic formula in the case $m=1$ see, e.g., \cite{mw65}.
The method used in \cite{mw65} can be also applied in the case
$m>0$.

Finally, a direct numerical computation of the multiple integrals
$W_d^{(m)}(z)$ is an awkward task. Fortunately, for $m>0$ and
$z\geq d$, it is reduced to the following integral of the Bessel
function $I_0(x)$:
\[
W_d^{(m)}(z)=\frac{1}{\Gamma(m)}\int_{0}^\infty\!dv\,v^{m-1}e^{-zv}I_0^d(v).
\]

\section{The main results.}

Usually thermodynamic properties are derived in the limit of an
infinitely large lattice. In our case the results are most
conveniently formulated in a continuum limit. We choose to use the
version of continuum limit where the limiting configurations are
random functions defined on the $d$-dimensional rectangle
$[0,1]^d$:
\[
\{x(\gamma)\}_{\gamma\in[0,1]^d}\equiv\{x(\gamma_1,\gamma_2,\ldots,
\gamma_d)\}_{\gamma_1,\gamma_2,\ldots,\gamma_d\in[0,1]}.
\]
For any $\gamma\in[0,1]^d$ the random variable $x(\gamma)$ is
defined as the following limit in distribution
\[
x(\gamma)\stackrel{d}{=}\lim_{n\to\infty}x_{([\gamma_1n],[\gamma_2n],\ldots,[\gamma_dn])},
\]
where $[y]$, is the integer part of $y$.

Thermodynamic random variables $x(\gamma)$ and $x(\delta)$ are
limits of the random sequences
$x_{([\gamma_1n],[\gamma_2n],\ldots,[\gamma_dn])}$ and
$x_{([\delta_1n],[\delta_2n],\ldots,[\delta_dn])}$ separated by a
distance of order $n$. Hence, in the continuum limit the random
variables $x(\gamma)$ and $x(\delta)$ with $\gamma\neq\delta$ are
independent due to the exponential/power-law decay of
thermodynamic correlations in the high/low temperature region.

Unless explicitly stated otherwise, in this paper we consider
dimensions $d\geq5$ and inverse temperatures $\beta>\beta_c$,
where $\beta_c^{-1}$ is the critical temperature of the spherical
model in external random field, see the paper by Pastur
\cite{p82}.

Denote $\varphi_{(1,1,\ldots,1)}$ the projection of the external
random field on the eigenvector $\mbox{\boldmath
$w$}^{(1,1,\ldots,1)}$ corresponding to the maximal eigenvalue of
the interaction matrix
\[
\varphi_{(1,1,\ldots,1)}=\sqrt{\frac{2}{n^{d-1}(n+1)}}\sum_{l\in
V_n}\sin\frac{\pi l_1}{n+1}\,h_l.
\]
Recall that the external field $\{h_l, l\in Z^d\}$ is a
realizations of the random field $\{\mbox{\boldmath $h$}_l, l\in
Z^d\}$ of independent normal random variables  with
$\mbox{\boldmath $E$}\mbox{\boldmath $h$}_l=0$, $\mbox{\boldmath
$E$}\mbox{\boldmath $h$}^2_l=h^2$, for any $l\in Z^d$. Everywhere
below we will use the notation ${\cal N}(a,b^2)$ to denote {\em
thermodynamic\/} normal random variables with mean $a$ and
variance $b^2$, which are independent from the external random
field. The symbol $q$ will be used to denote a realization of a
{\em non\/}-thermodynamic normal random variable $\mbox{\boldmath
$q$}$. The value of $\mbox{\boldmath $q$}$ is fixed  once we fix a
realization of the random field, and $\mbox{\boldmath $q$}$ is
always independent of ${\cal N}(a,b^2)$.

The main results of the paper can be stated as follows.
\begin{enumerate}
\item In the absence of the boundary field, $b=0$, the random
variables $x(\gamma)$ have normal distributions with the expected
values
\[
\langle
x(\gamma)\rangle=\mbox{sgn}\left[\varphi_{(1,1,\ldots,1)}\right]
\sin(\pi\gamma_1)\sqrt{\left(1-\frac{\beta_c}{\beta}\right)\frac{W_d^{(1)}(d)}{\beta_cJ}
}+q_\gamma,
%Z_\gamma(0,V_{\rm bulk}^2)
\]
and the variances
\[
\langle x^2(\gamma)\rangle-\langle x(\gamma)\rangle^2=
\frac{1}{2\beta J}W_d^{(1)}(d),
%=\frac{1}{2\beta J}\int_0^\infty\!dv\, e^{-vd}I_0^d(v).
%\int_{-\pi}^\pi\!\!\!\ldots\int_{-\pi}^\pi
%\frac{1}{d-\sum_{\nu=1}^d\cos\omega_\nu}\prod_{\nu=1}^d
%\frac{d\omega_\nu}{2\pi}.
\]
where $q_\gamma$ are independent realizations of zero-mean normal
random variables with the common variance
\[
\left(\frac{h}{2J}\right)^2W_d^{(2)}(d).
\]
\item For a fixed realization of the external random field, the
law of large numbers is valid for the normalized sums
\[
m_n\equiv\frac{1}{N}\sum_{j\in V_n}x_j,
\]
as $n\to\infty$. The convergence to the limiting value can be
summarized by the following asymptotic formula
\[
m_n\sim{\rm sgn}\left[\varphi_{(1,1,\ldots,1)}\right]\frac{2}{\pi}
\sqrt{\left(1-\frac{\beta_c}{\beta}\right)\frac{W_d^{(1)}(d)}{\beta_c
J}}+n^{2-d/2}q_n+
\]
\[
+\frac{n^{-d/4}}{\sqrt{|\varphi_{(1,1,\ldots,1)}|}}\,{\cal
N}_n\left(0,\frac{8}{\pi^2\beta}\sqrt{\left(1-
\frac{\beta_c}{\beta}\right)\frac{W_d^{(1)}(d)}{2\beta_c
J}}\right),
\]
where $q_n$ is a realization of a zero-mean normal random variable
with the variance
\[
2\,\frac{7\pi^2-69}{3\pi^6}\left(\frac{h}{2J}\right)^2.
\]
Hence, the magnetization $m_n$ is (only) {\em conditionally
self-averaging} with the exponents $\rho=\frac{1}{2}-\frac{2}{d}$
and $\tau=\frac{1}{4}$. \item For $b\neq0$ the random variables
$x(\gamma)$ have normal distributions with expected values
\[
\langle x(\gamma)\rangle=
\frac{b}{J}\frac{\cosh\left[(1-2\gamma_1)\sqrt{
\zeta_0}\right]}{\cosh\sqrt{\zeta_0}}+q_\gamma,
\]
and variances
\[
\langle x^2(\gamma)\rangle-\langle x(\gamma)\rangle^2=
\frac{1}{2\beta J}W_d^{(1)}(d),
%\int_{-\pi}^\pi\!\!\!\ldots\int_{-\pi}^\pi
%\frac{1}{d-\sum_{\nu=1}^d\cos\omega_\nu}\prod_{\nu=1}^d
%\frac{d\omega_\nu}{2\pi}.
\]
where $\zeta_0$ is a solution of Eq.\ (\ref{sp}), and $q_\gamma$
are independent realizations of zero-mean normal random variables
with the common variance
\[
\left(\frac{h}{2J}\right)^2W_d^{(2)}(d).
\]
\item For $b\neq0$, the law of large numbers is valid for the
normalized sums
\[
m_n\equiv\frac{1}{N}\sum_{j\in V_n}x_j,
\]
as $n\to\infty$. The convergence to the limiting value can be
summarized by the following asymptotic formula
\[
%\frac{1}{N}\sum_{j\in V_n}x_j
m_n\sim\frac{b}{J}\frac{\tanh\sqrt{\zeta_0}}
{\sqrt{\zeta_0}}+n^{2-d/2}q_n +n^{1-d/2}{\cal
N}_n\left(0,\frac{1}{4\beta
J\zeta_0}\left(1-\frac{\tanh\sqrt{\zeta_0}}
{\sqrt{\zeta_0}}\right)\right),
\]
where $q_n$ is a realization of a zero-mean normal random variable
with the variance (\ref{varq}). Hence, the magnetization $m_n$ is
self-averaging with the exponents $\rho=\frac{1}{2}-\frac{2}{d}$
and $\tau=\frac{1}{2}-\frac{1}{d}$.
\end{enumerate}

\section{The free energy.}
The calculation of free energy, expected values and correlation
functions for the spherical models is reduced, in a routine
fashion, to calculation of the large-$n$ asymptotics of an
integral. In this section we find the large-$n$ asymptotics for
the free energy
\[
f_n=-\frac{1}{\beta n^d}\ln\Theta_n.
\]
A particular attention will be paid to $O(n^{-2})$ asymptotics of
$f_n$, which, as it turns out, determines thermodynamic properties
of the model below the critical temperature.

The introduction of new integration variables $\{y_j\}_{j\in V_n}$
in Eq.\ (\ref{pf}) via the orthogonal transformation
\[
x_j=\sum_{k\in V_n}w_j^{(k)}y_k,\quad j\in V_n,
\]
where the eigenvectors $\{w_j^{(k)}\}_{j\in V_n}$ are given by
Eq.\ (\ref{evs}), diagonalises the interaction matrix. Therefore,
we obtain the following formula for the partition function
\[
\Theta_n=\int_{-\infty}^\infty\ldots\int_{-\infty}^\infty
e^{-\beta \widetilde H_n(y)}\mu_n(dy),
\]
where
\[
\widetilde H_n(y)=-J\sum_{k\in V_n}\lambda_k y_k^2-\sum_{k\in
V_n}\varphi_k y_k-b\sum_{k\in V_n}\alpha_k y_k,
\]
\[
\varphi_k=\sum_{j\in V_n}h_j w_j^{(k)},\quad\mbox{and}\quad
\alpha_k=\sum_{j\in B_n} w_j^{(k)}.
\]
Since the vectors $\{w_j^{(k)}\}_{j\in V_n}$, $k\in V_n$ are
orthonormal, the random variables $\mbox{ \boldmath
$\varphi$}_k=\sum_{j\in V_n}\mbox{ \boldmath $h$}_j w_j^{(k)}$,
are independent normal random variables with zero mean and
variance $h^2$. Therefore, we can treat the coefficients
$\varphi_k$, $k\in V_n$ as realizations of independent normal
random variables.

A direct calculation of the coefficients $\alpha_k$,
$k\equiv(k_1,k_2,\ldots,k_d)\in V_n$ (using only the formula for
the sum of a geometric series) yields
\[
\alpha_k=2n^{(d-1)/2}\sqrt{\frac{2}{n+1}}\,\delta(k_2,1)\ldots\delta(k_d,1)\times
\left\{\begin{array}{cl}
\sin{\displaystyle\frac{\pi k_1}{n+1}},&\mbox{ if $k_1$ is odd,}\vspace{1mm}\\
0,&\mbox{ if $k_1$ is even.}
\end{array}
\right.
\]

The integral representation for the delta function
\[
\delta\left(\sum_{j\in V_n}y_j^2-N\right)=\frac{1}{2\pi
i}\int_{-i\infty}^{+i\infty}\!ds\, \exp\left[s\left(N-\sum_{j\in
V_n}y_j^2\right)\right],
\]
in the {\em ``a priori''} measure allows one to perform
integration over the variables $y_j$, $j\in V_n$. However, we can
switch the order of integration over the variables $y_j$, $j\in
V_n$ and $s$ only after a shift of the integration contour for $s$
to the right. The shift should assure that the real part of the
quadratic form involving the variables $y_j$, $j\in V_n$ is
negatively defined. The switching of integration order,
integration over $y_j$, $j\in V_n$, and the introduction of a new
integration variable $z$ via $s=\beta J z$ yields
%\[
%\Theta_n=\frac{\beta J}{2\pi
%i}\int_{-i\infty+c}^{+i\infty+c}\!dz\, e^{\beta J z N}\prod_{k\in
%V_n}\left\{\int_{-\infty}^\infty\! dy\,\exp\left[-\beta
%J(z-\lambda_k)y^2+\beta(\varphi_k+b\alpha_k)y\right]\right\},
%\]
% On integrating over $y$ we obtain
\begin{equation}
\Theta_n=\frac{\beta J}{2\pi i}\left(\frac{\pi}{\beta
J}\right)^{N/2}\int_{-i\infty+c}^{+i\infty+c}\!dz\,
\exp\left[N\beta\Phi_n(z)\right], \label{pf1}
\end{equation}
where
\[
\Phi_n(z)=J z-\frac{1}{2\beta N}\sum_{k\in
V_n}\ln(z-\lambda_k)+\frac{1}{4JN}\sum_{k\in
V_n}\frac{(\varphi_k+b\alpha_k)^2}{z-\lambda_k},
\]
and $c>d$ is the shift of the integration contour mentioned above.

The large-$n$ asymptotics of the integral (\ref{pf1}) can be found
using the saddle-point method. The saddle point of the integrand
is a solution of the equation
\begin{equation}
\Phi_n'(z)=J -\frac{1}{2\beta N}\sum_{k\in
V_n}\frac{1}{z-\lambda_k}-\frac{1}{4JN}\sum_{k\in
V_n}\left(\frac{\varphi_k+b\alpha_k}{z-\lambda_k}\right)^2=0.
\label{sp1}
\end{equation}
For any $z>d$, as $n\to\infty$, the sequence of the derivatives
$\Phi_n'(z)$ converges, with probability 1, to
\[
\Phi'(z)= J-\frac{1}{2\beta}W_d^{(1)}(z)-
\frac{h^2}{4J}W_d^{(2)}(z),
\]
where the functions $W_d^{(m)}(z)$ are defined in Eq.\ (\ref{wf}).
%\begin{eqnarray}
%\Phi'(z)&=&\beta J-\frac{1}{2}\int_{-\pi}^{\pi}\ldots
%\int_{-\pi}^{\pi} \frac{1}{z-\sum_{\nu=1}^d \cos
%\omega_\nu}\prod_{\nu=1}^d
%\frac{d\omega_\nu}{2\pi}\nonumber\\&-&\frac{\beta}{4J}\int_{-\pi}^{\pi}\ldots
%\int_{-\pi}^{\pi}\frac{1}{\left(z-\sum_{\nu=1}^d \cos
%\omega_\nu\right)^2}\prod_{\nu=1}^d
%\frac{d\omega_\nu}{2\pi}.\nonumber
%\end{eqnarray}
The function $\Phi'(z)$ increases monotonically with $z$ on
$[d,\infty)$, and the location of its zeroes depends on the
dimension $d$ of the lattice. Namely, if $d\leq 4$, then the
function $\Phi'(z)$ has exactly one zero on the interval
$[d,\infty)$ at a point $z^*>d$, for any $\beta>0$. If $d\geq 5$
and the variance of the external field, $h^2$, is sufficiently
small, then there exists a critical value
\[
\beta_c
=\frac{1}{2J}\frac{W_d^{(1)}(d)}{1-\left(\frac{h}{2J}\right)^2W_d^{(2)}(d)}
\]
of the parameter $\beta$, see \cite{p82}. If $\beta\in(0,\beta_c)$
(the high-temperature regime), then the function $\Phi'(z)$ still
has exactly one zero on the interval $[d,\infty)$ at a point
$z^*>d$. While if $\beta > \beta_c$ (the low-temperature regime),
then the function $\Phi'(z)$ is strictly positive on the interval
$[d,\infty)$.

The application of the saddle-point method for the integral
(\ref{pf1}) is fairly straightforward when the saddle point $z^*$
is greater than $d$, see \cite{bk}. Therefore, in this paper we
consider only the low-temperature regime and $d\geq5$. When
$\beta\geq\beta_c$, the function $\Phi_n(z)$ still attains its
minimum on the interval $(\lambda_{\rm max},\infty)$ at a point
$z_n^*>\lambda_{\rm max}$, where $\lambda_{\rm
max}=d-1+\cos\frac{\pi}{n+1}$ is the maximum eigenvalue of the
interaction matrix $\widehat T$. However, the sequence of saddle
points $z_n^*$ approaches the branch point of the integrand at
$z=\lambda_{\rm max}$, and the application of the saddle-point
method becomes a bit more tricky.

To be able to apply the saddle-point method we have to find a
change of variables $z=\lambda_{\rm max}+\zeta n^{-\gamma}$, such
that the sequence of rescaled saddle-points
$\zeta^*_n=(z_n^*-\lambda_{\rm max})n^\gamma$ converges to a
positive limit $\zeta^*>0$ as $n\to \infty$. Then, the application
of the saddle-point method for the integral over $\zeta$ becomes
straightforward again. Note that the above search for a proper
change of variables has an important physical meaning ---
$n^{\gamma/2}/\sqrt{\zeta^*}$ is the correlation length of the
model.

In order to find the proper value of $\gamma$ we have to analyse
the sums in Eq.\ (\ref{sp1}). The large-$n$ asymptotics of the sum
\[
\Sigma_1(z)\equiv\frac{1}{N}\sum_{k\in V_n} \frac{1}{z-\lambda_k},
\]
when $z=\lambda_{\rm max}+\zeta n^{-\gamma}$ and $\zeta>0$,
follows from Eqs.\ (\ref{stoi0}) and (\ref{stoi}). Namely, as
$n\to\infty$,
\[
\Sigma_1(\lambda_{\rm max}+\zeta n^{-\gamma})=\frac{1}{\zeta
n^{d-\gamma}}+W_d^{(1)}(d)+O(n^{-\min(\gamma,1)}).
%
%\int_{-\pi}^{\pi}\ldots \int_{-\pi}^{\pi} \frac{1}{\lambda_{\rm
%max}+\zeta n^{-\gamma}-\sum_{\nu=1}^d \cos
%\omega_\nu}\prod_{\nu=1}^d \frac{d\omega_\nu}{2\pi},
\]

To find the large-$n$ asymptotics of the sum
\[
\Sigma_2(z)\equiv\frac{1}{N}\sum_{k\in
V_n}\frac{\varphi_k^2}{(z-\lambda_k)^2}
\]
when $z=\lambda_{\rm max}+\zeta n^{-\gamma}$, we have to use the
law of large numbers. First, we take out the term corresponding to
$k=(1,1,\ldots,1)$ and rearrange the sum as follows
\begin{eqnarray}
\Sigma_2(\lambda_{\rm max}+\zeta
n^{-\gamma})&=&\frac{\varphi_{(1,1,\ldots,1)}^2}{\zeta^2
n^{d-2\gamma}}+\frac{1}{N}{\sum_{k\in
V_n}}'\frac{h^2}{(\lambda_{\rm
max}+\zeta n^{-\gamma}-\lambda_k)^2}\nonumber\\
&+&\frac{1}{N}{\sum_{k\in
V_n}}'\frac{\varphi_k^2-h^2}{(\lambda_{\rm max}+\zeta
n^{-\gamma}-\lambda_k)^2}. \nonumber
\end{eqnarray}
For $\zeta\geq0$, Eqs.\ (\ref{stoi0}) and (\ref{stoim}) yield as
$n\to\infty$
\[
\frac{1}{N}{\sum_{k\in V_n}}'\frac{1}{(\lambda_{\rm max}+\zeta
n^{-\gamma}-\lambda_k)^2}=W_d^{(2)}(d)+o(1).
\]

Let $\{\xi_{j,n}\}_{j=1,n=1}^{n\;\;\;\;\;\infty}$ be a triangular
array of independent random variables with zero expected values.
The condition
\[
\sum_{j=1}^n\mbox{\boldmath $E$}|\xi_{j,n}|^s\to0,\quad\mbox{for
some }s\in(1,2],
\]
as $n\to\infty$, is sufficient for the validity of the law of
large numbers
\[
\sum_{j=1}^n\xi_{j,n}\to 0,\quad\mbox{in probability},
\]
see, e.g.,\ \cite{b00}. Therefore Eqs.\ (\ref{wf}), (\ref{stoi0}),
and (\ref{stoim}) imply
\[
\frac{1}{N}{\sum_{k\in V_n}}'\frac{\varphi_k^2-h^2}{(\lambda_{\rm
max}+\zeta n^{-\gamma}-\lambda_k)^2}\to0,\quad\mbox{in
probability},
\]
as $n\to\infty$, if $d>4$, and $\zeta\geq0$. Summarizing the above
we obtain
\[
\Sigma_2(\lambda_{\rm max}+\zeta
n^{-\gamma})=\frac{\varphi_{(1,1,\ldots,1)}^2}{\zeta^2
n^{d-2\gamma}}+h^2W_d^{(2)}(d)+o(1),
\]
as $n\to\infty$.

The sum
\[
\Sigma_3(z)\equiv\frac{1}{N}\sum_{k\in
V_n}\frac{\varphi_k\alpha_k}{(z-\lambda_k)^2},
\]
with $z=\lambda_{\rm max}+\zeta n^{-\gamma}$, is a realisation of
a normal random variable with zero mean and the variance
\[
\sigma_n^2(\zeta)=\frac{1}{N^2}\sum_{k\in
V_n}\frac{h^2\alpha_k^2}{(\lambda_{\rm max}+\zeta
n^{-\gamma}-\lambda_k)^4}.
\]
It is possible to find a relatively simple expression for the
variance
\[
\sigma_n^2(\zeta)=\frac{2h^2}{n^{d+1}(n+1)}\sum_{k=1}^n\frac{(1+(-1)^{k+1})^2
\sin^2\frac{\pi k}{n+1}}{\left(\cos\frac{\pi}{n+1}+\zeta
n^{-\gamma}-\cos\frac{\pi k}{n+1}\right)^4}.
\]
First, note the identity, see \cite{p94},
\begin{equation}
\frac{1}{N}\sum_{k\in V_n}\frac{\alpha_k^2}{z-\lambda_k}=
\frac{4x(z)}{n}\frac{x^{n-1}(z)+1}{x^{n+1}(z)+1}, \label{sum1}
\end{equation}
where
\[
x(z)=1+z-d+\sqrt{(z-d)(2+z-d)}.
\]
On differentiating Eq.\ (\ref{sum1}) over $z$ three times we
obtain
\[
\frac{1}{N^2}\sum_{k\in
V_n}\frac{\alpha_k^2}{(z-\lambda_k)^4}=\frac{1}{n^{d+1}}\frac{2}{(z-d)(2+z-d)}
\left[\frac{1+z-d}{[(z-d)(2+z-d)]^{3/2}}\frac{x^{n+1}(z)-1}{x^{n+1}(z)+1}\right.
\]
\[
\left.-\frac{4(n+1)^3x^{2(n+1)}(z)}{\left(x^{n+1}(z)+1\right)^4}+
\frac{2(n+1)x^{(n+1)}(z)}{\left(x^{n+1}(z)+1\right)^2}
\left(\frac{n(n+2)}{3}-\frac{1}{(z-d)(2+z-d)}\right)\right].
\]
Hence, if $\gamma\in(0,2)$, then
\[
\sigma_n^2(\zeta)\sim
{\displaystyle\frac{1}{n^{d+1-5\gamma/2}}\frac{2h^2}{(2\zeta)^{5/2}}},
\]
as $n\to\infty$, while if $\gamma=2$, then $\lambda_{\rm
max}+\zeta n^{-\gamma}\sim d-\frac{1}{2}\pi^2n^{-2}+\zeta n^{-2}$,
and $\sigma_n^2(\zeta)\sim
n^{4-d}h^2\,t(\zeta-{\textstyle\frac{1}{2}}\pi^2)$, where
\[
t(\zeta)=\displaystyle{\frac{1}{\zeta}\left( \frac{1}{
(2\zeta)^{3/2}}\tanh\sqrt{{\textstyle\frac{1}{2}}\zeta}-
\frac{1}{4\cosh^4\sqrt{\frac{1}{2}\zeta}}+
\frac{2\zeta-3}{12\zeta\cosh^2\sqrt{\frac{1}{2}\zeta}}\right)}.
\]
The function $t(\zeta)$ (and similar functions below) has only a
removable singularity at $\zeta=0$, and the analytic continuation
is to be used for negative values of $\zeta$. Thus
\[
\Sigma_3(\lambda_{\rm max}+\zeta n^{-\gamma})\sim
O\left(n^{5\gamma/4-(d+1)/2}\right)
\]
does not produce
a non-vanishing contribution to the saddle-point equation if
$\gamma\leq2$.

It is also possible to obtain a simple formula for the sum
\[
\Sigma_4(z)=\frac{1}{N}\sum_{k\in
V_n}\frac{\alpha_k^2}{(z-\lambda_k)^2}
\]
by differentiating Eq.\ (\ref{sum1}) over $z$. The differentiation
yields
\[
\Sigma_4(z)=\frac{8}{n}\left[\frac{x^{n-1}(z)-1}{x^{n+1}(z)+1}\frac{x^{2}(z)}{x^{2}(z)-1}+
\frac{(n+1)x^{n+1}(z)}{(x^{n+1}(z)+1)^2}\right].
\]
On replacing $z$ by $\lambda_{\rm max}+\zeta n^{-\gamma}$ we
obtain
\[
\Sigma_4(\lambda_{\rm max}+\zeta n^{-\gamma})\sim\left\{
\begin{array}{cl}
{\displaystyle\frac{4n^{-1+\gamma/2}}{\sqrt{\frac{1}{2}\zeta}}},&\rm{
if
}\ \gamma\in(0,2);\vspace{2mm}\\
{\displaystyle
\frac{2\tanh\sqrt{\frac{1}{2}(\zeta-\frac{1}{2}\pi^2)}}{
\sqrt{\frac{1}{2}(\zeta-\frac{1}{2}\pi^2)}}+
\frac{2}{\cosh^2\sqrt{\frac{1}{2}(\zeta-\frac{1}{2}\pi^2)}}},&\rm{
if }\ \gamma=2.
\end{array} \right.
\]
Thus the sum $\Sigma_4(z)$ is dominant among the four sums
$\Sigma_l(z)$, $l=1,2,3,4$ (if $b\neq0$), in the sense that it is
$\Sigma_4(z)$ that controls the location of the saddle point
$z^*_n$ in the low-temperature region. Indeed, the sum
$\Sigma_4(z)$ produces a non-vanishing contribution to the
saddle-point equation already in the scale $z=\lambda_{\rm
max}+\zeta/n^2$. Moreover, the extra contribution produced by
$\Sigma_4(z)$ prevents the rescaled saddle-point $\zeta_n^*$
approaching the branch-point at $\zeta=0$, where the remaining
sums could, potentially, yield non-vanishing contributions to the
saddle-point equation.

On introduction of the new integration variable $\zeta$ in Eq.\
(\ref{pf1}) via $z=\lambda_{\rm max}+\zeta n^{-2}$ we obtain
\[
\Theta_n=\frac{\beta J}{2n^2\pi i}\left(\frac{\pi}{\beta
J}\right)^{N/2}\int_{-i\infty+\zeta_0}^{+i\infty+\zeta_0}\!d\zeta\,
\exp\left[N\beta\Phi_n(\lambda_{\rm max}+\zeta n^{-2})\right].
\]
The saddle-point of the integrand is
$\zeta^*=2\zeta_0+\frac{1}{2}\pi^2$, where $\zeta_0$ is a solution
of the equation
\begin{equation}
1-\frac{1}{2\beta J}W_d^{(1)}(d)-
\left(\frac{h}{2J}\right)^2W_d^{(2)}(d)=2\left(\frac{b}{2J}\right)^2
\left(\frac{\tanh\sqrt{\zeta_0}} {\sqrt{\zeta_0}}+
\frac{1}{\cosh^2\sqrt{\zeta_0}}\right). \label{sp}
\end{equation}
Application of the saddle-point method yields
\[
-f_n=\frac{1}{\beta n^d}\ln\Theta_n=
%\frac{\beta
%J}{n^{2+d/2}}\left(\frac{\pi}{\beta
%J}\right)^{N/2}\sqrt{\frac{1}{2\pi\Phi''_n(d+\zeta^*
%n^{-2})}}\times
\frac{1}{2\beta}\ln\frac{\pi}{\beta J}
+\Phi(d)+n^{-1}\phi_1+n^{-2}\phi_2\left(2\zeta_0\right)+o(n^{-2}),
\]
as $n\to\infty$, where
\begin{eqnarray}
&&\Phi(d)=Jd-\frac{1}{2\beta}L_d(d)+\frac{h^2}{4J}W_d^{(1)}(d),\nonumber\\
&&L_d(z)=\int_{-\pi}^{\pi}\!\ldots \int_{-\pi}^{\pi}
\ln\left(z-\sum_{\nu=1}^d \cos \omega_\nu\right)\prod_{l=1}^d
\frac{d\omega_l}{2\pi},\nonumber\\
&&\phi_1=\frac{1}{4\beta}\Delta L_{d}(d)-\frac{h^2}{8J}\Delta
W_{d}^{(1)}(d)+\frac{b^2}{J},\nonumber\\
&&\phi_2(\zeta)=\left( J-\frac{1}{2\beta}W_d^{(1)}(d)-\frac{
h^2}{4J}W_d^{(2)}(d)\right)\zeta-\frac{
b^2}{J}\sqrt{2\zeta}\tanh\sqrt{{\textstyle\frac{1}{2}}\zeta}.
\nonumber
\end{eqnarray}

The function $\Phi(z)$ determines the thermodynamics of the model
in the high-temperature region. The term $\phi_1$ appears because
of to the lack of periodicity in one of the dimensions. The
function $\phi_2(\zeta)$ is responsible for the thermodynamic
properties of the model on the low-temperature critical line.

\section{Individual distributions.}

To find the individual distributions of the random variables
$\{x_j,j\in V_n\}$ we calculate the corresponding characteristic
functions
\[
\kappa_j(t)=\langle\exp(itx_j)\rangle.
\]
The saddle-point method described in the previous section yields
the following large-$n$ asymptotics
\begin{equation}
\kappa_j(t)\sim \exp\left[-\frac{t^2}{4\beta J}\sum_{k\in
V_n}\frac{\left(w_j^{(k)}\right)^2}{z_n^*-\lambda_k}+\frac{it}{2
J}\sum_{k\in V_n}\frac{\varphi_k w_j^{(k)}+b\alpha_k
w_j^{(k)}}{z_n^* -\lambda_k}\right]. \label{cfid}
\end{equation}
Therefore, for large values of $n$, the individual distributions
of the random variables $\{x_j,j\in V_n\}$ are nearly normal with
mean values
\begin{equation}
\mu_j=\frac{1}{2 J}\sum_{k\in V_n}\frac{\varphi_k
w_j^{(k)}+b\alpha_k w_j^{(k)}}{z_n^* -\lambda_k},
\label{ev1}
\end{equation}
and variances
\[
\sigma_j^2=\frac{1}{2\beta J}\sum_{k\in
V_n}\frac{\left(w_j^{(k)}\right)^2}{z_n^*-\lambda_k}.
\]
On substitution $z_n^*=d+\zeta^* n^{-2}$ one obtains
\[
\sigma_j^2\to\frac{1}{2\beta
J}\int_{-\pi}^\pi\!\ldots\int_{-\pi}^\pi \frac{1-\cos(j_1
\omega_1)}{d-\sum_{\nu=1}^d\cos
\omega_\nu}\prod_{l=1}^d\frac{d\omega_l}{2\pi},
\]
as $n\to\infty$. Thus, in the low-temperature region and in the
presence of the boundary conditions, $b\neq0$, the variances of
the thermodynamic random variables $x_j$ are not affected by the
random field $\{h_l,\ l\in Z^d\}$. As $j_1$ increases,
\[
\int_{-\pi}^\pi\!\ldots\int_{-\pi}^\pi \frac{\cos(j_1
\omega_1)}{d-\sum_{\nu=1}^d\cos
\omega_\nu}\prod_{l=1}^d\frac{d\omega_l}{2\pi}\sim
\frac{\Gamma(d/2-1)}{2\pi^{d/2}j_1^{d-2}}\to0.
\]
Hence, only random variables near the boundary have variances
noticeably different from the bulk value
\[
\sigma_{\rm bulk}^2\equiv\frac{1}{2\beta J}W_d^{(1)}(d).
%\int_{-\pi}^\pi\!\ldots\int_{-\pi}^\pi
%\frac{1}{d-\sum_{\nu=1}^d\cos
%\omega_\nu}\prod_{l=1}^d\frac{d\omega_l}{2\pi}
%=\frac{1}{2\beta J}\int_{0}^\infty\!dv\,e^{-vd}I_0^d(v),
\]
%where $I_0(v)$ is a Bessel function.

The first half of the sum in Eq.\ (\ref{ev1})
\[
q_j\equiv\frac{1}{2 J}\sum_{k\in V_n}\frac{\varphi_k
w_j^{(k)}}{z_n^*-\lambda_k},
\]
describes the shift in the expected value of $x_j$ due to the
external random field. It is a realization of a normal random
variable with zero mean and variance
\[
V_j^2\equiv\left(\frac{h}{2 J}\right)^2\sum_{k\in V_n}\left(\frac{
w_j^{(k)}}{z_n^*-\lambda_k}\right)^2.
\]
As $n\to\infty$ the variance $V_j^2$ tends to
\[
\left(\frac{h}{2 J}\right)^2\int_{-\pi}^\pi\!\ldots\int_{-\pi}^\pi
\frac{1-\cos(j_1 \omega_1)}{(d-\sum_{\nu=1}^d\cos
\omega_\nu)^2}\prod_{l=1}^d\frac{d\omega_l}{2\pi}.
\]
For $d>4$ we have
\[
\int_{-\pi}^\pi\!\ldots\int_{-\pi}^\pi \frac{\cos(j_1
\omega_1)}{(d-\sum_{\nu=1}^d\cos
\omega_\nu)^2}\prod_{l=1}^d\frac{d\omega_l}{2\pi}\sim
\frac{\Gamma(d/2-2)}{4\pi^{d/2}j_1^{d-4}},
\]
as $j_1\to\infty$. Hence, the variance $V_j^2$ also approaches its
bulk value
\begin{equation}
V_{\rm bulk}^2\equiv\left(\frac{h}{2 J}\right)^2W_d^{(2)}(d),
%\int_{-\pi}^\pi\!\ldots\int_{-\pi}^\pi
%\frac{1}{(d-\sum_{\nu=1}^d\cos
%\omega_\nu)^2}\prod_{l=1}^d\frac{d\omega_l}{2\pi}
%=\frac{h^2}{4 J^2}\int_{0}^\infty\!dv\,ve^{-vd}I_0^d(v),
\label{vbulk}
\end{equation}
as we move away from the boundary.

The second half of the sum in Eq.\ (\ref{ev1}),
\[
\mu_j^{\rm bc}\equiv\frac{b}{2 J}\sum_{k\in V_n}\frac{\alpha_k
w_j^{(k)}}{z_n^*-\lambda_k},
\]
is the shift in the expected value of the thermodynamic random
variables $x_j$ due to the influence of the boundary conditions.
An application of the ``contour summation" technique, see
\cite{p94}, yields the following simple formula
\[
\mu_j^{\rm bc}=\frac{b}{J}\frac{x^{n+1-j_1}(z^*_n)
+x^{j_1}(z^*_n)}{x^{n+1}(z^*_n)+1}.
\]
The large-$n$ limit of $\mu_j^{\rm bc}$ depends on the location of
the node $j\equiv(j_1,j_2,\ldots,j_d)$. Assuming $j_1\sim \gamma_1
n$ as $n\to\infty$, we obtain (recall that $z_n^*=\lambda_{\rm
max}+\zeta^* n^{-2}\sim d+2\zeta_0n^{-2}$ in the low-temperature
region, see Eq.\ (\ref{sp}))
\[
\lim_{n\to\infty}\mu_j^{\rm
bc}=\frac{b}{J}\frac{\cosh\left[(1-2\gamma_1)\sqrt{\zeta_0}\right]}{
\cosh\sqrt{\zeta_0}}\equiv\mu^{\rm bc}(\gamma_1).
\]

The characteristic function of an arbitrary pair $(x_j,x_l)$ is
given by
\[
\kappa_{j,l}(t,s)=\langle\exp(itx_j+isx_l)\rangle\sim\kappa_{j}(t)
\kappa_{l}(s)\exp\left[-\frac{ts}{2\beta J}\sum_{k\in
V_n}\frac{w_j^{(k)}w_l^{(k)}}{z_n^*-\lambda_k}\right],
\]
as $n\to\infty$. Hence, for large values of $n$, the joint
distribution of $x_j$ and $x_l$ is nearly normal with the
covariance
\[
{\rm cov}(x_j,x_l)\sim\frac{1}{2\beta J}\sum_{k\in
V_n}\frac{w_j^{(k)}w_l^{(k)}}{z_n^*-\lambda_k}.
\]
Since $z_n^*=\lambda_{\rm max}+\zeta^*n^{-2}$, we have (ignoring
thin layers near the boundaries)
\[
\lim_{n\to\infty}{\rm cov}(x_j,x_l)=\frac{1}{2\beta
J}\int_{-\pi}^{\pi}\ldots\int_{-\pi}^{\pi}
\frac{\exp\left[i\sum_{\nu=1}^d (j_\nu-l_\nu)\omega_\nu\right]}
{d-\sum_{\nu=1}^d \cos\omega_\nu}\prod_{\nu=1}^d
\frac{d\omega_\nu}{2\pi}.
\]
Thus, the covariance ${\rm cov}(x_j,x_l)$ shows the usual, for the
critical line of the ordinary spherical model, power-law decay
with the distance $r_{j,l}^2\equiv\sum_{\nu=1}^d(j_\nu-l_\nu)^2$
between the nodes $j$ and $l$. Indeed, using Eq.\ (\ref{gf}) we
obtain
\begin{equation}
{\rm cov}(x_j,x_l)\sim\frac{\Gamma(d/2-1)}{4\beta
J\pi^{d/2}r_{j,l}^{d-2}},
\label{dcf}
\end{equation}
if $1\ll r_{j,l}\ll n$.

Summarizing, we conclude that the structure of random variables
$\{x_j,j\in V_n\}$ is fairly simple. Ignoring thin layers near
boundaries, we have in the limit $n\to\infty$
\[
x_j=q_j+{\cal N}_j(\mu^{\rm bc}(\gamma_1),\sigma_{\rm bulk}^2),
\]
where $q_j$ is a realization of a (non-thermodynamic) normal
random variable with zero mean and the variance $V_{\rm bulk}^2$
and ${\cal N}_j(a,b^2)$ is a thermodynamic normal random variable
with the mean $a$ and the variance $b^2$, see Fig.\ 1.

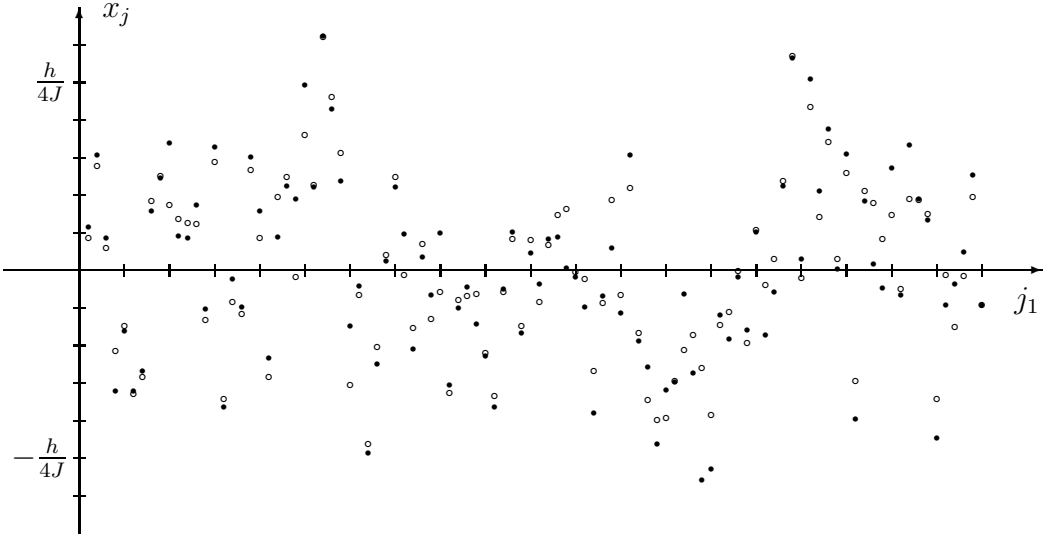
\begin{figure}
\setlength{\unitlength}{1mm}
\begin{picture}(150,70)(0,0)
\put(0,35){\vector(1,0){138}} \put(10,0){\vector(0,1){70}}
\put(15,69){\makebox(0,0){$x_j$}}
%\put(10,50){\makebox(0,0){$\rule{1.5mm}{0.15mm}$}}
\put(4,59){$\frac{h}{4J}$} \put(1,9){$-\frac{h}{4J}$}
\put(136,31){\makebox(0,0){$j_1$}}
\multiput(10,5)(0,5){13}{\makebox(0,0){$\rule{1.5mm}{0.15mm}$}}
\multiput(10,35)(6,0){21}{\makebox(0,0){$\rule{0.15mm}{1.5mm}$}}
%\put(24,-5){\makebox(0,0){$t_{300}$}}
%\put(38,-5){\makebox(0,0){$t_{600}$}}
%\put(52,-5){\makebox(0,0){$t_{900}$}}
%\put(66,-5){\makebox(0,0){$t_{1200}$}}
%\put(80,-5){\makebox(0,0){$t_{1500}$}}
%\put(94,-5){\makebox(0,0){$t_{1800}$}}
%\put(108,-5){\makebox(0,0){$t_{2100}$}}
\put(11.2,40.8068){\circle*{0.7}}
\put(11.2,39.3435){\circle{0.7}}
\put(12.4,50.3098){\circle*{0.7}}
\put(12.4,48.9246){\circle{0.7}}
\put(13.6,39.3522){\circle*{0.7}}
\put(13.6,37.9493){\circle{0.7}}
\put(14.8,18.8873){\circle*{0.7}}
\put(14.8,24.2457){\circle{0.7}}
\put(16.,26.8787){\circle*{0.7}}
\put(16.,27.6585){\circle{0.7}}
\put(17.2,18.9862){\circle*{0.7}}
\put(17.2,18.5401){\circle{0.7}}
\put(18.4,21.5615){\circle*{0.7}}
\put(18.4,20.8233){\circle{0.7}}
\put(19.6,42.8635){\circle*{0.7}}
\put(19.6,44.2845){\circle{0.7}}
\put(20.8,47.2879){\circle*{0.7}}
\put(20.8,47.5879){\circle{0.7}}
\put(22.,51.975){\circle*{0.7}}
\put(22.,43.7417){\circle{0.7}}
\put(23.2,39.5396){\circle*{0.7}}
\put(23.2,41.8303){\circle{0.7}}
\put(24.4,39.3377){\circle*{0.7}}
\put(24.4,41.3113){\circle{0.7}}
\put(25.6,43.6608){\circle*{0.7}}
\put(25.6,41.1006){\circle{0.7}}
\put(26.8,29.9081){\circle*{0.7}}
\put(26.8,28.443){\circle{0.7}}
\put(28.,51.4336){\circle*{0.7}}
\put(28.,49.3893){\circle{0.7}}
\put(29.2,16.8504){\circle*{0.7}}
\put(29.2,17.87){\circle{0.7}}
\put(30.4,33.9182){\circle*{0.7}}
\put(30.4,30.7377){\circle{0.7}}
\put(31.6,30.0567){\circle*{0.7}}
\put(31.6,29.1879){\circle{0.7}}
\put(32.8,50.0421){\circle*{0.7}}
\put(32.8,48.414){\circle{0.7}}
\put(34.,42.8769){\circle*{0.7}}
\put(34.,39.2853){\circle{0.7}}
\put(35.2,23.3916){\circle*{0.7}}
\put(35.2,20.8414){\circle{0.7}}
\put(36.4,39.516){\circle*{0.7}}
\put(36.4,44.7057){\circle{0.7}}
\put(37.6,46.2269){\circle*{0.7}}
\put(37.6,47.4782){\circle{0.7}}
\put(38.8,44.5205){\circle*{0.7}}
\put(38.8,34.1942){\circle{0.7}}
\put(40.,59.6361){\circle*{0.7}}
\put(40.,52.9973){\circle{0.7}}
\put(41.2,46.1228){\circle*{0.7}}
\put(41.2,46.4153){\circle{0.7}}
\put(42.4,66.225){\circle*{0.7}}
\put(42.4,66.1033){\circle{0.7}}
\put(43.6,56.4308){\circle*{0.7}}
\put(43.6,58.0914){\circle{0.7}}
\put(44.8,46.847){\circle*{0.7}}
\put(44.8,50.6755){\circle{0.7}}
\put(46.,27.5834){\circle*{0.7}}
\put(46.,19.7002){\circle{0.7}}
\put(47.2,32.9397){\circle*{0.7}}
\put(47.2,31.6956){\circle{0.7}}
\put(48.4,10.7159){\circle*{0.7}}
\put(48.4,11.9183){\circle{0.7}}
\put(49.6,22.5272){\circle*{0.7}}
\put(49.6,24.7833){\circle{0.7}}
\put(50.8,36.2282){\circle*{0.7}}
\put(50.8,37.0134){\circle{0.7}}
\put(52.,46.1122){\circle*{0.7}}
\put(52.,47.4373){\circle{0.7}}
\put(53.2,39.8591){\circle*{0.7}}
\put(53.2,34.4332){\circle{0.7}}
\put(54.4,24.589){\circle*{0.7}}
\put(54.4,27.4039){\circle{0.7}}
\put(55.6,36.8165){\circle*{0.7}}
\put(55.6,38.4604){\circle{0.7}}
\put(56.8,31.7739){\circle*{0.7}}
\put(56.8,28.5333){\circle{0.7}}
\put(58.,39.9196){\circle*{0.7}}
\put(58.,32.17){\circle{0.7}}
\put(59.2,19.7331){\circle*{0.7}}
\put(59.2,18.7302){\circle{0.7}}
\put(60.4,29.9852){\circle*{0.7}}
\put(60.4,31.1099){\circle{0.7}}
\put(61.6,32.7814){\circle*{0.7}}
\put(61.6,31.5433){\circle{0.7}}
\put(62.8,27.8069){\circle*{0.7}}
\put(62.8,31.9229){\circle{0.7}}
\put(64.,23.6775){\circle*{0.7}}
\put(64.,24.022){\circle{0.7}}
\put(65.2,16.7637){\circle*{0.7}}
\put(65.2,18.2614){\circle{0.7}}
\put(66.4,32.52){\circle*{0.7}}
\put(66.4,32.1048){\circle{0.7}}
\put(67.6,40.0519){\circle*{0.7}}
\put(67.6,39.169){\circle{0.7}}
\put(68.8,26.6255){\circle*{0.7}}
\put(68.8,27.6064){\circle{0.7}}
\put(70.,37.3437){\circle*{0.7}}
\put(70.,39.0504){\circle{0.7}}
\put(71.2,33.2634){\circle*{0.7}}
\put(71.2,30.7415){\circle{0.7}}
\put(72.4,39.2226){\circle*{0.7}}
\put(72.4,38.3053){\circle{0.7}}
\put(73.6,39.5126){\circle*{0.7}}
\put(73.6,42.4493){\circle{0.7}}
\put(74.8,35.2877){\circle*{0.7}}
\put(74.8,43.1876){\circle{0.7}}
\put(76.,34.086){\circle*{0.7}}
\put(76.,34.8524){\circle{0.7}}
\put(77.2,30.0788){\circle*{0.7}}
\put(77.2,33.9096){\circle{0.7}}
\put(78.4,15.9978){\circle*{0.7}}
\put(78.4,21.5797){\circle{0.7}}
\put(79.6,31.5932){\circle*{0.7}}
\put(79.6,30.6734){\circle{0.7}}
\put(80.8,37.953){\circle*{0.7}}
\put(80.8,44.3623){\circle{0.7}}
\put(82.,29.3199){\circle*{0.7}}
\put(82.,31.6871){\circle{0.7}}
\put(83.2,50.3875){\circle*{0.7}}
\put(83.2,45.9518){\circle{0.7}}
\put(84.4,25.5976){\circle*{0.7}}
\put(84.4,26.6205){\circle{0.7}}
\put(85.6,22.1503){\circle*{0.7}}
\put(85.6,17.7619){\circle{0.7}}
\put(86.8,11.8646){\circle*{0.7}}
\put(86.8,15.0497){\circle{0.7}}
\put(88.,19.0839){\circle*{0.7}}
\put(88.,15.3904){\circle{0.7}}
\put(89.2,20.1193){\circle*{0.7}}
\put(89.2,20.2459){\circle{0.7}}
\put(90.4,31.916){\circle*{0.7}}
\put(90.4,24.436){\circle{0.7}}
\put(91.6,21.3082){\circle*{0.7}}
\put(91.6,26.4784){\circle{0.7}}
\put(92.8,7.07718){\circle*{0.7}}
\put(92.8,22.0491){\circle{0.7}}
\put(94.,8.61294){\circle*{0.7}}
\put(94.,15.7852){\circle{0.7}}
\put(95.2,29.1053){\circle*{0.7}}
\put(95.2,27.7837){\circle{0.7}}
\put(96.4,25.9054){\circle*{0.7}}
\put(96.4,29.4454){\circle{0.7}}
\put(97.6,34.1277){\circle*{0.7}}
\put(97.6,34.9486){\circle{0.7}}
\put(98.8,27.1227){\circle*{0.7}}
\put(98.8,25.3689){\circle{0.7}}
\put(100.,40.105){\circle*{0.7}}
\put(100.,40.4481){\circle{0.7}}
\put(101.2,26.4057){\circle*{0.7}}
\put(101.2,33.0637){\circle{0.7}}
\put(102.4,32.1084){\circle*{0.7}}
\put(102.4,36.592){\circle{0.7}}
\put(103.6,46.1934){\circle*{0.7}}
\put(103.6,46.9538){\circle{0.7}}
\put(104.8,63.2953){\circle*{0.7}}
\put(104.8,63.4515){\circle{0.7}}
\put(106.,36.5597){\circle*{0.7}}
\put(106.,33.9802){\circle{0.7}}
\put(107.2,60.508){\circle*{0.7}}
\put(107.2,56.7001){\circle{0.7}}
\put(108.4,45.5784){\circle*{0.7}}
\put(108.4,42.1312){\circle{0.7}}
\put(109.6,53.8628){\circle*{0.7}}
\put(109.6,52.0783){\circle{0.7}}
\put(110.8,35.2328){\circle*{0.7}}
\put(110.8,36.4886){\circle{0.7}}
\put(112.,50.4464){\circle*{0.7}}
\put(112.,47.9286){\circle{0.7}}
\put(113.2,15.2148){\circle*{0.7}}
\put(113.2,20.339){\circle{0.7}}
\put(114.4,44.1987){\circle*{0.7}}
\put(114.4,45.6204){\circle{0.7}}
\put(115.6,35.8562){\circle*{0.7}}
\put(115.6,44.012){\circle{0.7}}
\put(116.8,32.6499){\circle*{0.7}}
\put(116.8,39.1215){\circle{0.7}}
\put(118.,48.5837){\circle*{0.7}}
\put(118.,42.3674){\circle{0.7}}
\put(119.2,31.7084){\circle*{0.7}}
\put(119.2,32.5768){\circle{0.7}}
\put(120.4,51.7299){\circle*{0.7}}
\put(120.4,44.5376){\circle{0.7}}
\put(121.6,44.4782){\circle*{0.7}}
\put(121.6,44.4194){\circle{0.7}}
\put(122.8,41.7113){\circle*{0.7}}
\put(122.8,42.544){\circle{0.7}}
\put(124.,12.7405){\circle*{0.7}}
\put(124.,17.8418){\circle{0.7}}
\put(125.2,30.4625){\circle*{0.7}}
\put(125.2,34.4389){\circle{0.7}}
\put(126.4,33.1702){\circle*{0.7}}
\put(126.4,27.5122){\circle{0.7}}
\put(127.6,37.4662){\circle*{0.7}}
\put(127.6,34.2592){\circle{0.7}}
\put(128.8,47.6338){\circle*{0.7}}
\put(128.8,44.754){\circle{0.7}}
\put(130.,30.3747){\circle*{0.7}}
\put(130.,30.4559){\circle{0.7}}
\end{picture}
\caption{A line of a 5-D realization of thermodynamic random
variables $\{\mbox{\boldmath $x$}_j, j\in V_n\}$ (discs) for
$\beta>\beta_c$. The picture also contains the corresponding
realization of the random field $\{\mbox{\boldmath $q$}_j, j\in
V_n\}$ (circles) driving the random variables at very low
temperatures.}
\end{figure}

In the presence of the boundary conditions, apart from the global
influence through the saddle point $\zeta^*$, the external random
field $\{h_j,j\in V_n\}$ produces only additive contributions
(random shifts) $q_j$ to the thermodynamic random variables
$\{x_j,j\in V_n\}$. The properties of the (non-thermodynamic)
random variables $\{\mbox{\boldmath $q$}_j,j\in V_n\}$ generating
the shifts are fairly interesting. At the critical temperature
$\beta_c$ the random field $\{\mbox{\boldmath $q$}_j,j\in V_n\}$
undergoes a transition into a phase with long-range correlations,
see Fig.\ 2.

\begin{figure}
\setlength{\unitlength}{1mm}
\begin{picture}(150,70)(0,0)
\put(0,35){\vector(1,0){138}} \put(10,0){\vector(0,1){70}}
\put(15,69){\makebox(0,0){$q_j$}}
%\put(10,50){\makebox(0,0){$\rule{1.5mm}{0.15mm}$}}
\put(4,59){$\frac{h}{4J}$} \put(1,9){$-\frac{h}{4J}$}
\put(136,31){\makebox(0,0){$j_1$}}
\multiput(10,5)(0,5){13}{\makebox(0,0){$\rule{1.5mm}{0.15mm}$}}
\multiput(10,35)(6,0){21}{\makebox(0,0){$\rule{0.15mm}{1.5mm}$}}
%\put(24,-5){\makebox(0,0){$t_{300}$}}
%\put(38,-5){\makebox(0,0){$t_{600}$}}
%\put(52,-5){\makebox(0,0){$t_{900}$}}
%\put(66,-5){\makebox(0,0){$t_{1200}$}}
%\put(80,-5){\makebox(0,0){$t_{1500}$}}
%\put(94,-5){\makebox(0,0){$t_{1800}$}}
%\put(108,-5){\makebox(0,0){$t_{2100}$}}
\put(11.2,40.4293){\circle*{0.7}}%\put(11.2,40.4293){\circle{0.7}}
\put(12.4,52.4058){\circle*{0.7}}%\put(12.4,51.6265){\circle{0.7}}
\put(13.6,38.6867){\circle*{0.7}}%\put(13.6,31.4372){\circle{0.7}}
\put(14.8,21.5571){\circle*{0.7}}%\put(14.8,18.103){\circle{0.7}}
\put(16.,25.8231){\circle*{0.7}}%\put(16.,29.9199){\circle{0.7}}
\put(17.2,14.4251){\circle*{0.7}}%\put(17.2,16.1175){\circle{0.7}}
\put(18.4,17.2791){\circle*{0.7}}%\put(18.4,24.4158){\circle{0.7}}
\put(19.6,46.6056){\circle*{0.7}}%\put(19.6,55.786){\circle{0.7}}
\put(20.8,50.7348){\circle*{0.7}}%\put(20.8,48.3533){\circle{0.7}}
\put(22.,45.9271){\circle*{0.7}}%\put(22.,40.9478){\circle{0.7}}
\put(23.2,43.5378){\circle*{0.7}}%\put(23.2,39.7207){\circle{0.7}}
\put(24.4,42.8891){\circle*{0.7}}%\put(24.4,39.6637){\circle{0.7}}
\put(25.6,42.6257){\circle*{0.7}}%\put(25.6,39.4972){\circle{0.7}}
\put(26.8,26.8038){\circle*{0.7}}%\put(26.8,22.2){\circle{0.7}}
\put(28.,52.9867){\circle*{0.7}}%\put(28.,57.4504){\circle{0.7}}
\put(29.2,13.5875){\circle*{0.7}}%\put(29.2,3.60745){\circle{0.7}}
\put(30.4,29.6721){\circle*{0.7}}%\put(30.4,37.5404){\circle{0.7}}
\put(31.6,27.7349){\circle*{0.7}}%\put(31.6,29.1149){\circle{0.7}}
\put(32.8,51.7675){\circle*{0.7}}%\put(32.8,56.5834){\circle{0.7}}
\put(34.,40.3567){\circle*{0.7}}%\put(34.,34.1047){\circle{0.7}}
\put(35.2,17.3018){\circle*{0.7}}%\put(35.2,13.2377){\circle{0.7}}
\put(36.4,47.1321){\circle*{0.7}}%\put(36.4,55.3303){\circle{0.7}}
\put(37.6,50.5978){\circle*{0.7}}%\put(37.6,47.0433){\circle{0.7}}
\put(38.8,33.9927){\circle*{0.7}}%\put(38.8,27.2673){\circle{0.7}}
\put(40.,57.4966){\circle*{0.7}}%\put(40.,59.511){\circle{0.7}}
\put(41.2,49.2691){\circle*{0.7}}%\put(41.2,40.6502){\circle{0.7}}
\put(42.4,73.8792){\circle*{0.7}}%\put(42.4,70.6946){\circle{0.7}}
\put(43.6,63.8642){\circle*{0.7}}%\put(43.6,49.0807){\circle{0.7}}
\put(44.8,54.5944){\circle*{0.7}}%\put(44.8,42.4483){\circle{0.7}}
\put(46.,15.8752){\circle*{0.7}}%\put(46.,3.43358){\circle{0.7}}
\put(47.2,30.8694){\circle*{0.7}}%\put(47.2,36.0425){\circle{0.7}}
\put(48.4,6.14787){\circle*{0.7}}%\put(48.4,3.53201){\circle{0.7}}
\put(49.6,22.2291){\circle*{0.7}}%\put(49.6,32.131){\circle{0.7}}
\put(50.8,37.5168){\circle*{0.7}}%\put(50.8,42.9195){\circle{0.7}}
\put(52.,50.5466){\circle*{0.7}}%\put(52.,51.2214){\circle{0.7}}
\put(53.2,34.2915){\circle*{0.7}}%\put(53.2,27.793){\circle{0.7}}
\put(54.4,25.5049){\circle*{0.7}}%\put(54.4,24.4943){\circle{0.7}}
\put(55.6,39.3255){\circle*{0.7}}%\put(55.6,43.2313){\circle{0.7}}
\put(56.8,26.9167){\circle*{0.7}}%\put(56.8,24.1366){\circle{0.7}}
\put(58.,31.4625){\circle*{0.7}}%\put(58.,34.3053){\circle{0.7}}
\put(59.2,14.6627){\circle*{0.7}}%\put(59.2,14.1292){\circle{0.7}}
\put(60.4,30.1373){\circle*{0.7}}%\put(60.4,38.2524){\circle{0.7}}
\put(61.6,30.6791){\circle*{0.7}}%\put(61.6,32.76){\circle{0.7}}
\put(62.8,31.1537){\circle*{0.7}}%\put(62.8,33.1707){\circle{0.7}}
\put(64.,21.2775){\circle*{0.7}}%\put(64.,22.0891){\circle{0.7}}
\put(65.2,14.0768){\circle*{0.7}}%\put(65.2,18.3134){\circle{0.7}}
\put(66.4,31.381){\circle*{0.7}}%\put(66.4,40.4848){\circle{0.7}}
\put(67.6,40.2113){\circle*{0.7}}%\put(67.6,43.2851){\circle{0.7}}
\put(68.8,25.758){\circle*{0.7}}%\put(68.8,23.7206){\circle{0.7}}
\put(70.,40.063){\circle*{0.7}}%\put(70.,45.0869){\circle{0.7}}
\put(71.2,29.6769){\circle*{0.7}}%\put(71.2,27.706){\circle{0.7}}
\put(72.4,39.1316){\circle*{0.7}}%\put(72.4,42.2501){\circle{0.7}}
\put(73.6,44.3116){\circle*{0.7}}%\put(73.6,43.9201){\circle{0.7}}
\put(74.8,45.2344){\circle*{0.7}}%\put(74.8,42.6251){\circle{0.7}}
\put(76.,34.8155){\circle*{0.7}}%\put(76.,30.5268){\circle{0.7}}
\put(77.2,33.637){\circle*{0.7}}%\put(77.2,33.3316){\circle{0.7}}
\put(78.4,18.2246){\circle*{0.7}}%\put(78.4,16.9701){\circle{0.7}}
\put(79.6,29.5918){\circle*{0.7}}%\put(79.6,36.0193){\circle{0.7}}
\put(80.8,46.7029){\circle*{0.7}}%\put(80.8,50.4074){\circle{0.7}}
\put(82.,30.8589){\circle*{0.7}}%\put(82.,26.0569){\circle{0.7}}
\put(83.2,48.6897){\circle*{0.7}}%\put(83.2,51.8787){\circle{0.7}}
\put(84.4,24.5256){\circle*{0.7}}%\put(84.4,17.8803){\circle{0.7}}
\put(85.6,13.4523){\circle*{0.7}}%\put(85.6,15.6377){\circle{0.7}}
\put(86.8,10.0621){\circle*{0.7}}%\put(86.8,16.7481){\circle{0.7}}
\put(88.,10.488){\circle*{0.7}}%\put(88.,19.2356){\circle{0.7}}
\put(89.2,16.5574){\circle*{0.7}}%\put(89.2,26.2887){\circle{0.7}}
\put(90.4,21.795){\circle*{0.7}}%\put(90.4,29.9131){\circle{0.7}}
\put(91.6,24.348){\circle*{0.7}}%\put(91.6,30.6979){\circle{0.7}}
\put(92.8,18.8113){\circle*{0.7}}%\put(92.8,23.5516){\circle{0.7}}
\put(94.,10.9815){\circle*{0.7}}%\put(94.,17.2205){\circle{0.7}}
\put(95.2,25.9796){\circle*{0.7}}%\put(95.2,36.9954){\circle{0.7}}
\put(96.4,28.0567){\circle*{0.7}}%\put(96.4,33.2573){\circle{0.7}}
\put(97.6,34.9357){\circle*{0.7}}%\put(97.6,39.8682){\circle{0.7}}
\put(98.8,22.9611){\circle*{0.7}}%\put(98.8,23.6558){\circle{0.7}}
\put(100.,41.8102){\circle*{0.7}}%\put(100.,49.0255){\circle{0.7}}
\put(101.2,32.5796){\circle*{0.7}}%\put(101.2,30.9978){\circle{0.7}}
\put(102.4,36.99){\circle*{0.7}}%\put(102.4,39.4001){\circle{0.7}}
\put(103.6,49.9422){\circle*{0.7}}%\put(103.6,51.5343){\circle{0.7}}
\put(104.8,70.5643){\circle*{0.7}}%\put(104.8,68.5272){\circle{0.7}}
\put(106.,33.7253){\circle*{0.7}}%\put(106.,19.079){\circle{0.7}}
\put(107.2,62.1251){\circle*{0.7}}%\put(107.2,64.7476){\circle{0.7}}
\put(108.4,43.914){\circle*{0.7}}%\put(108.4,32.8537){\circle{0.7}}
\put(109.6,56.3479){\circle*{0.7}}%\put(109.6,53.7989){\circle{0.7}}
\put(110.8,36.8607){\circle*{0.7}}%\put(110.8,27.1019){\circle{0.7}}
\put(112.,51.1608){\circle*{0.7}}%\put(112.,50.731){\circle{0.7}}
\put(113.2,16.6737){\circle*{0.7}}%\put(113.2,7.0933){\circle{0.7}}
\put(114.4,48.2756){\circle*{0.7}}%\put(114.4,56.1247){\circle{0.7}}
\put(115.6,46.265){\circle*{0.7}}%\put(115.6,41.2107){\circle{0.7}}
\put(116.8,40.1518){\circle*{0.7}}%\put(116.8,35.3736){\circle{0.7}}
\put(118.,44.2092){\circle*{0.7}}%\put(118.,42.1212){\circle{0.7}}
\put(119.2,31.971){\circle*{0.7}}%\put(119.2,26.9993){\circle{0.7}}
\put(120.4,46.922){\circle*{0.7}}%\put(120.4,48.4655){\circle{0.7}}
\put(121.6,46.7743){\circle*{0.7}}%\put(121.6,42.2124){\circle{0.7}}
\put(122.8,44.4301){\circle*{0.7}}%\put(122.8,39.6207){\circle{0.7}}
\put(124.,13.5523){\circle*{0.7}}%\put(124.,6.58633){\circle{0.7}}
\put(125.2,34.2986){\circle*{0.7}}%\put(125.2,42.1877){\circle{0.7}}
\put(126.4,25.6403){\circle*{0.7}}%\put(126.4,24.6311){\circle{0.7}}
\put(127.6,34.0741){\circle*{0.7}}%\put(127.6,37.7774){\circle{0.7}}
\put(128.8,47.1925){\circle*{0.7}}%\put(128.8,48.7758){\circle{0.7}}
\put(130.,29.3198){\circle*{0.7}}%\put(130.,23.7343){\circle{0.7}}
\end{picture}
\caption{A line of a 5-D realization of the random field
$\{\mbox{\boldmath $q$}_j, j\in V_n\}$ for $\beta>\beta_c$.  In
comparison with independent random variables the field
$\{\mbox{\boldmath $q$}_j, j\in V_n\}$ has a substantial inertia
--- positive/negative values tend to be surrounded by positive/negative values.}
%\caption{A line of a 5-D realization of the random field
%$\{\mbox{\boldmath $q$}_j, j\in V_n\}$ (discs) for
%$\beta>\beta_c$. To outline the correlations, the picture also
%contains a realization of independent normal random variables with
%the same variance (circles) used to generate the realization of
%the random field. A close look at the picture reveals that in
%comparison to the independent random variables the field
%$\{\mbox{\boldmath $q$}_j, j\in V_n\}$ has a substantial inertia
%--- long-range positive correlations.}
\end{figure}
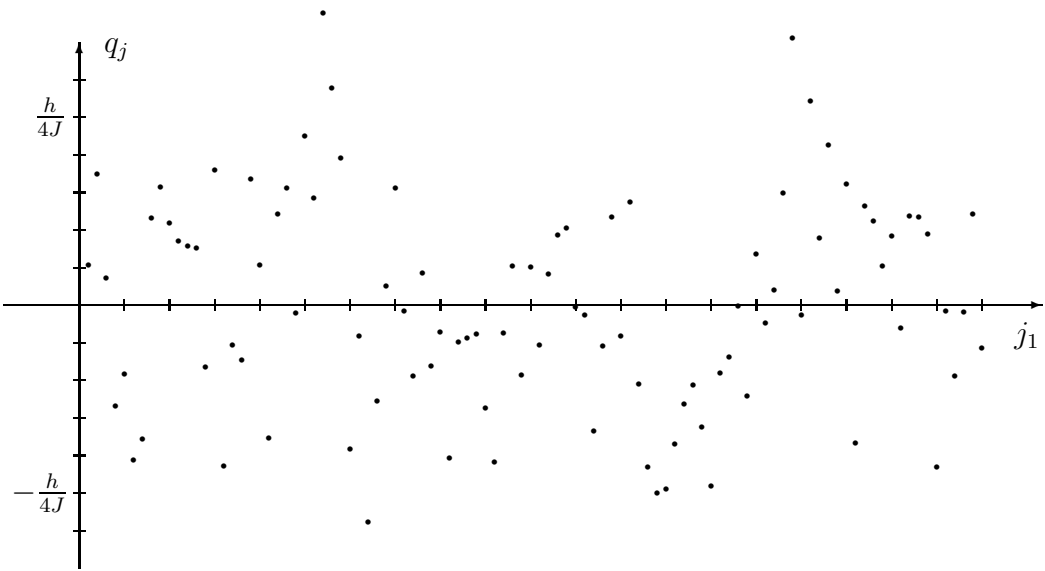

Indeed the covariances of the random variables $\{\mbox{\boldmath
$q$}_j,j\in V_n\}$ are given by
\[
{\rm cov}(\mbox{\boldmath $q$}_j,\mbox{\boldmath
$q$}_l)=\left(\frac{h}{2J}\right)^2\sum_{k\in
V_n}\frac{w_j^{(k)}w_l^{(k)}}{(z_n^*-\lambda_k)^2}.
\]
Passing to the limit $n\to\infty$ we obtain
\[
\lim_{n\to\infty}{\rm cov}(\mbox{\boldmath $q$}_j,\mbox{\boldmath
$q$}_l)=\left(\frac{h}{2
J}\right)^2\int_{-\pi}^{\pi}\ldots\int_{-\pi}^{\pi}
\frac{\exp\left[i\sum_{\nu=1}^d (j_\nu-l_\nu)\omega_\nu\right]}
{\left(z^*-\sum_{\nu=1}^d \cos\omega_\nu\right)^2}\prod_{\nu=1}^d
\frac{d\omega_\nu}{2\pi}.
\]
If $\beta<\beta_c$, then $z^*>d$ and the above integral decays
exponentially with the distance $r_{j,l}$ between the nodes $j$
and $l$. If $\beta\geq\beta_c$, then $z^*=d$ and Eq.\ (\ref{gf})
yields the power-law decay
\[
{\rm cov}(\mbox{\boldmath $q$}_j,\mbox{\boldmath $q$}_l)\sim
\frac{h^2\Gamma(d/2-2)}{16J^2\pi^{d/2}r_{j,l}^{d-4}}.
\]
%if $1\ll r_{j,l}\ll n$.
Note that the correlations of the random field $\{\mbox{\boldmath
$q$}_j,j\in V_n\}$ decay noticeably slower than the correlations
of the thermodynamic random variables, see Eq.\ (\ref{dcf}). This
slow decay of the covariances is the reason for the dominance of
the random-field fluctuations over the thermodynamic fluctuations.

%Kosterlitz-Thouless transition at $\beta_c$. Namely, we have an
%exponential decay of correlations for $\beta<\beta_c$, and a
%power-law decay  for $\beta>\beta_c$.
%Moreover, the transition is
%not accompanied by symmetry breaking. Indeed the distribution of
%the random variables $\{\mbox{\boldmath $q$}_j\}_{j\in V_n}$ is
%neither shifted from zero, nor split into peaks --- the two most
%common scenarios for a transition accompanied by a symmetry
%breaking.

\section{Macroscopic observables.}

Our aim in this section is to establish the law of large numbers
for the normalized sums (magnetization)
\begin{equation}
m_n\equiv\frac{1}{N}\sum_{j\in V_n}x_j, \label{tm}
\end{equation}
and to study fluctuations (the central limit theorem) of these
sums around the limiting value. The corresponding characteristic
functions are given by
\[
\kappa_n(t)=\left\langle\exp\left(\frac{it}{N}\sum_{j\in
V_n}x_j\right)\right\rangle.
\]
The large-$n$ asymptotics of $\kappa_n(t)$ is calculated using the
technique of the previous section. The saddle-point method yields
\begin{equation}
\kappa_n(t)\sim \exp\left(-\frac{t^2}{4\beta JN^2}\sum_{k\in
V_n}\frac{\eta_k^2}{z_n^*-\lambda_k}+\frac{it}{2JN}\sum_{k\in
V_n}\frac{\varphi_k \eta_k+b\alpha_k\eta_k}{z_n^*
-\lambda_k}\right), \label{cft}
\end{equation}
where (see Eq.\ (\ref{evs}))
\[
\eta_k=\sum_{j\in
V_n}w_j^{(k)}=n^{(d-1)/2}\sqrt{\frac{2}{n+1}}\frac{1-(-1)^{k_1}}{2}\frac{
\sin{\frac{\pi k_1}{n+1}}}{1-\cos{\frac{\pi
k_1}{n+1}}}\,\delta(k_2,1)\ldots\delta(k_d,1),
\]
for $k\equiv(k_1,k_2,\ldots,k_d)\in V_n$.

Thus, for large values of $n$, the distribution of the
magnetization (\ref{tm}) is approximately normal with the mean
value
\begin{equation}
\mu_n=\frac{1}{2 JN}\sum_{k\in V_n}\frac{\varphi_k
\eta_k+b\alpha_k\eta_k}{z_n^* -\lambda_k}. \label{ev2}
\end{equation}
The sum
\[
\frac{1}{2 JN}\sum_{k\in V_n}\frac{\varphi_k \eta_k}{z_n^*
-\lambda_k}=\frac{n^{-(d+1)/2}}{4J}\sqrt{\frac{2}{n+1}}
\sum_{l=1}^n\frac{\varphi_{(l,1,1,\ldots,1)}[1-(-1)^l]}{1+z_n^*-d-\cos\frac{\pi
l}{n+1}}\frac{\sin\frac{\pi l}{n+1}}{1-\cos\frac{\pi l}{n+1}},
\]
is the shift in the expected value of the magnetization (\ref{tm})
caused by the external random field. It is a realization of a
normal random variable with zero mean and the variance
\[
S_n^2\equiv\frac{h^2n^{-d-1}}{4J^2(n+1)}
\sum_{l=1}^n\frac{1-(-1)^l}{\left(1+z_n^*-d-\cos\frac{\pi
l}{n+1}\right)^2}\frac{\sin^2\frac{\pi
l}{n+1}}{\left(1-\cos\frac{\pi l}{n+1}\right)^2}.
\]
On calculating the sum over $l$, see \cite{p94}, we obtain
\begin{equation}
S_n^2=\frac{h^2n^{-d-1}}{2J^2(z_n^*-d)^2}\left[\frac{n}{2}-\frac{
2x^{n+1}(z_n^*)+x^{n}(z_n^*)-x(z_n^*)-2}
{\left(x^{n+1}(z_n^*)+1\right)\left(x(z_n^*)-x^{-1}(z_n^*)\right)}+
\frac{(n+1)x^{n+1}(z_n^*)}
{\left(x^{n+1}(z_n^*)+1\right)^2}\right]. \label{vex}
\end{equation}
In the low-temperature region we have $z_n^*=\lambda_{\rm
max}+\zeta^* n^{-2}\sim d+2\zeta_0n^{-2}$, see Eq.\ (\ref{sp}),
therefore
\begin{equation}
S_n^2\sim \left(\frac{h}{2J}\right)^2
\frac{n^{4-d}}{4\zeta_0^2}\left(
1-\frac{3\tanh\sqrt{\zeta_0}}{2\sqrt{\zeta_0}}+
\frac{1}{2\cosh^2\sqrt{\zeta_0}}\right), \label{vex0}
\end{equation}
as $n\to\infty$.

The sum
\[
\frac{b}{2 JN}\sum_{k\in V_n}\frac{\alpha_k \eta_k}{z_n^*
-\lambda_k}=\frac{b}{Jn}\frac{2x(z_n^*)}{x(z_n^*)-1}\frac{x^n(z_n^*)-1}{x^{n+1}(z_n^*)+1},
\]
is the shift in the expected value of the magnetization (\ref{tm})
caused by the boundary conditions. On substitution
$z_n^*=\lambda_{\rm max}+\zeta^* n^{-2}$ one obtains
\begin{equation}
\frac{b}{2 JN}\sum_{k\in V_n}\frac{\alpha_k \eta_k}{z_n^*
-\lambda_k}\sim\frac{b}{J}\frac{\tanh\sqrt{\zeta_0}}
{\sqrt{\zeta_0}}, \label{expb}
\end{equation}
as $n\to\infty$.

Let's now look at the variance of the magnetization. According to
Eq.\ (\ref{cft}) it is given by
\[
\sigma^2\equiv\frac{1}{2\beta JN^2}\sum_{k\in
V_n}\frac{\eta_k^2}{z_n^*-\lambda_k}=\frac{n^{-d-1}}{2\beta
J(n+1)}\sum_{l=1}^n\frac{1-(-1)^l}{1+z_n^*-d-\cos\frac{\pi
l}{n+1}}\frac{\sin^2\frac{\pi l}{n+1}}{\left(1-\cos\frac{\pi
l}{n+1}\right)^2}.
\]
The remaining sum over $l$ can be calculated exactly, and we
obtain the following expression for the variance
\begin{equation}
\sigma^2=\frac{n^{-d-1}}{2\beta
J(z_n^*-d)}\left[n-\frac{2x(z_n^*)\left(x^n(z_n^*)-1\right)}
{\left(x(z_n^*)-1\right)\left(x^{n+1}(z_n^*)+1\right)}\right].
\label{vart}
\end{equation}
On substitution of $z_n^*=\lambda_{\rm max}+\zeta^* n^{-2}$ for
the saddle-point one obtains
\begin{equation}
\sigma^2\sim\frac{n^{2-d}}{4\beta
J\zeta_0}\left(1-\frac{\tanh\sqrt{\zeta_0}}
{\sqrt{\zeta_0}}\right), \label{vex1}
\end{equation}
as $n\to\infty$.

Summarizing the above we obtain the following expression for the
magnetization
\begin{equation}
m_n\sim\frac{b}{J}\frac{\tanh\sqrt{\zeta_0}}
{\sqrt{\zeta_0}}+n^{2-d/2}q_n+n^{1-d/2}{\cal
N}_n\left(0,\frac{1}{4\beta
J\zeta_0}\left(1-\frac{\tanh\sqrt{\zeta_0}}
{\sqrt{\zeta_0}}\right)\right), \label{mgnb}
\end{equation}
where $q_n$ is a realization of a zero-mean normal random variable
with the variance
\begin{equation}
\left(\frac{h}{2J}\right)^2 \frac{1}{4\zeta_0^2}\left(
1-\frac{3\tanh\sqrt{\zeta_0}}{2\sqrt{\zeta_0}}+
\frac{1}{2\cosh^2\sqrt{\zeta_0}}\right), \label{varq}
\end{equation}
and ${\cal N}_n(\mu,v^2)$ is a thermodynamic normal random
variable with mean $\mu$ and variance $v^2$. Therefore, the
magnetization of the spherical model is self-averaging (for
$b\neq0$) with the exponents $\rho=\frac{1}{2}-\frac{2}{d}$ and
$\tau=\frac{1}{2}-\frac{1}{d}$. The limiting magnetization
$m=\lim_{n\to\infty}m_n$ as a function of the boundary field $b$
is shown on Fig.\ 3.

\begin{figure}
\setlength{\unitlength}{1mm}
\begin{picture}(150,70)(0,0)
\put(10,35){\vector(1,0){130}} \put(75,0){\vector(0,1){70}}
\put(80,69){\makebox(0,0){$m$}}
\put(71,61.5){$1$}\put(77,5.5){$-1$}
\put(139,39){\makebox(0,0){$b/J$}}
\multiput(75,7)(0,7){9}{\makebox(0,0){$\rule{1.5mm}{0.15mm}$}}
\multiput(15,35)(10,0){13}{\makebox(0,0){$\rule{0.15mm}{1.5mm}$}}
\put(24,31){\makebox(0,0){$-5$}} \put(125,31){\makebox(0,0){$5$}}
\input Magnet
\end{picture}
\caption{The infinite-lattice magnetization
$m=\lim_{n\to\infty}m_n$ as a function of the normalized boundary
field $b/J$, for $d=5$, $\beta J=2$, and $h/J=0.5$. The left/right
limits at $b=0$ are given by ${\displaystyle
\mp\frac{2\sqrt{2}}{\pi}\sqrt{1-\frac{\beta_c}{\beta}}}$.}
\end{figure}
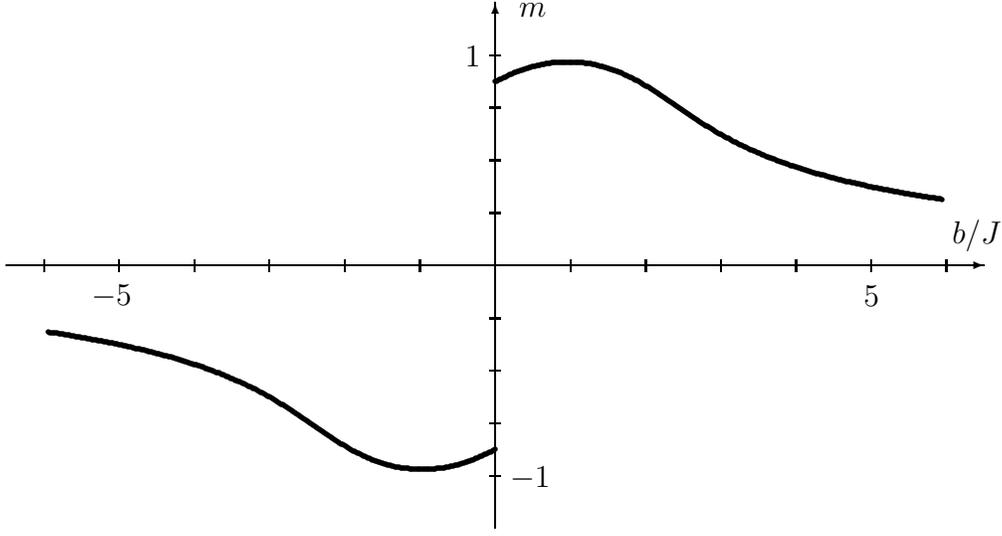

\section{The distributions for zero boundary field.}
As it is clear from previous sections a non-zero boundary field
dominates over the (zero-mean) random field in the low-temperature
regime. Therefore in this section we consider the case of zero
boundary field.

If $b=0$, then the saddle-point equation for the integral
(\ref{pf1}) is given by
\begin{equation}
\Phi_n'(z)\equiv J -\frac{1}{2\beta N}\sum_{k\in
V_n}\frac{1}{z-\lambda_k}-\frac{1}{4JN}\sum_{k\in
V_n}\left(\frac{\varphi_k}{z-\lambda_k}\right)^2=0. \label{sp2}
\end{equation}
Again, the saddle-point $z_n^*$ drifts towards the branch-point of
the integrand in the scale where the terms corresponding to
$k=(1,1,\ldots,1)$ produce a non-vanishing contribution to
$\Phi_n'(z)$. As is obvious from Eq.\ (\ref{sp2}), that happens in
the scale $z=\lambda_{(1,1,\ldots,1)}+\zeta n^{-d/2}$. The
distance from the saddle point
$z_n^*=\lambda_{(1,1,\ldots,1)}+\zeta^* n^{-d/2}$ to the
eigenvalues $\lambda_k$ with $k\neq(1,1,\ldots,1)$ is at least of
the order $O(n^{-2})$. Therefore there are no additional
non-vanishing contribution to the saddle-point equation from those
eigenvalues.

In the scale $z=\lambda_{(1,1,\ldots,1)}+\zeta n^{-d/2}$ we obtain
the following saddle-point equation in the limit $n\to\infty$
\[
1-\frac{1}{2\beta J}W_d^{(1)}(d)-\left(\frac{
h}{2J}\right)^2W_d^{(2)}(d)-\frac{1}{4J^2}\frac{\varphi^2_{(1,1,\ldots,1)}}{\zeta^2}=0.
\]
The positive solution of the above equation is given by
\[
\zeta^*=\frac{|\varphi_{(1,1,\ldots,1)}|}{2J
\sqrt{\left(1-\frac{\beta_c}{\beta}\right)\frac{1}{2J\beta_c}W_d^{(1)}(d)}}.
\]
The location of the saddle-point $z_n^*$, as $n\to\infty$, is
given by
\[
z_n^*\sim d-1+\cos\frac{\pi}{n+1}+\zeta^*n^{-d/2}.
\]

Evaluation of the characteristic function (\ref{cfid}) at $z_n^*$
shows that the thermodynamic variables $x_j$ have normal
distributions with the expected values
\[
\mu_j=\frac{1}{2J}\frac{\varphi_{(1,1,\ldots,1)}
w_j^{(1,1,\ldots,1)}}{z_n^*-\lambda_{(1,1,\ldots,1)}}+\frac{1}{2J}\sum_{k\in
V_n\setminus(1,1,\ldots,1)}\frac{\varphi_k
w_j^{(k)}}{z_n^*-\lambda_k},
\]
and variances
\[
\sigma_j^2=\frac{1}{2\beta J}\sum_{k\in V_n}\frac{
\left(w_j^{(k)}\right)^2}{z_n^*-\lambda_k}.
\]

Assuming that $j=(j_1,j_2,\ldots,j_d)$, and that for
$k=1,2,\ldots,d$ we have $j_k\sim\gamma_k n$ with
$\gamma_k\in(0,1)$, we obtain
\begin{equation}
\lim_{n\to\infty}\mu_j={\rm
sgn}\left[\varphi_{(1,1,\ldots,1)}\right]
\sin(\pi\gamma_1)\sqrt{\left(1-\frac{\beta_c}{\beta}
\right)\frac{W_d^{(1)}(d)}{\beta_c J}}+q_\gamma, \label{ev3}
\end{equation}
where $\gamma\equiv(\gamma_1,\gamma_2,\ldots,\gamma_d)$, and
$q_\gamma$ are realizations of independent zero-mean normal random
variables with the variance $V^2_{\rm bulk}$ given by Eq.\
(\ref{vbulk}). An important feature of Eq.\ (\ref{ev3}) is the
term ${\rm sgn\,}\varphi_{(1,1,\ldots,1)}$ common to all expected
values $\mu_j$. This term is the reason for the absence of
conventional self-averaging for the normalized sums
(magnetization)
\begin{equation}
m_n\equiv\frac{1}{N}\sum_{j\in V_n}x_j. \label{ns}
\end{equation}

On substitution of the saddle point
$z_n^*=\lambda_{(1,1,\ldots,1)}+\zeta^* n^{-d/2}$ in Eq.\
(\ref{cft}) we see that, as $n\to\infty$, the distribution of the
magnetization (\ref{ns}) is asymptotically normal with the
expected value
\begin{eqnarray}
\mu_n &=&\frac{1}{2 JN}\sum_{k\in V_n}\frac{\varphi_k
\eta_k}{z_n^* -\lambda_k}
%=\frac{1}{2JN}\frac{\varphi_{(1,1,\ldots,1)}
%\eta_{(1,1,\ldots,1)}}{z_n^*-\lambda_{(1,1,\ldots,1)}}+\frac{1}{2
%JN}\sum_{k\in V_n\setminus(1,1,\ldots,1)}\frac{\varphi_k
%\eta_k}{z_n^* -\lambda_k}
\sim{\rm sgn}\left[\varphi_{(1,1,\ldots,1)}\right]\frac{2}{\pi}
\sqrt{
\left(1-\frac{\beta_c}{\beta}\right)\frac{W_d^{(1)}(d)}{\beta_c
J}}\nonumber\\
&+&\frac{1}{2 JN}\sum_{k\in
V_n\setminus(1,1,\ldots,1)}\frac{\varphi_k \eta_k}{z_n^*
-\lambda_k}.\nonumber
\end{eqnarray}
On subtracting the contribution of the maximum eigenvalue from
Eq.\ (\ref{vex}) one finds that the remaining sum over $k$ is a
realization of a normal random variable with zero mean and the
variance
\[
S_n^2\sim
2\frac{7\pi^2-69}{3\pi^6}\left(\frac{h}{2J}\right)^2n^{4-d},
\]
as $n\to\infty$.

On substitution of the saddle point
$z_n^*=\lambda_{(1,1,\ldots,1)}+\zeta^* n^{-d/2}$ in Eq.\
(\ref{vart}) we find that the thermodynamic variance of the
normalized sums (\ref{ns}) is given by
\[
\sigma^2_n\sim\frac{8}{\pi^2}\frac{1}
{|\varphi_{(1,1,\ldots,1)}|\beta n^{d/2}}
\sqrt{\left(1-\frac{\beta_c}{\beta}\right)
\frac{W_d^{(1)}(d)}{2\beta_c
J}}.
\]
as $n\to\infty$.

Summarizing the above we obtain the following expression for the
magnetization
\[
m_n={\rm sgn}\left[\varphi_{(1,1,\ldots,1)}\right]\frac{2}{\pi}
\sqrt{\left(1-\frac{\beta_c}{\beta}\right)\frac{W_d^{(1)}(d)}{\beta_c
J}}+n^{2-d/2}q_n+
\]
\begin{equation}
+\frac{n^{-d/4}}{\sqrt{|\varphi_{(1,1,\ldots,1)}|}}\,{\cal
N}_n\left(0,\frac{8}{\pi^2\beta}\sqrt{\left(1-
\frac{\beta_c}{\beta}\right)\frac{W_d^{(1)}(d)}{2\beta_c
J}}\right), \label{mgn}
\end{equation}
where $q_n$ is a realization of a zero-mean normal random variable
with the variance
\[
2\frac{7\pi^2-69}{3\pi^6}\left(\frac{h}{2J}\right)^2,
\]
and ${\cal N}_n(0,v^2)$ is a zero-mean thermodynamic normal random
variable with variance $v^2$. Thus, in the absence of the boundary
field, the magnetization of the spherical model is conditionally
self-averaging with the exponents $\rho=\frac{1}{2}-\frac{2}{d}$
and $\tau=\frac{1}{4}$.

\section{Discussion and concluding remarks.}

It was shown in the paper \cite{apz92} that there are problems
with almost sure convergence of Gibbs states for the random-field
Curie-Weiss model in the infinite-volume limit. In fact, below the
critical temperature, the limits of thermodynamic averages
$\langle s_j\rangle_N$ do not exist, almost surely, as the volume
$N$ tends to infinity. A possible solution of the convergence
problem was also proposed: it is necessary to consider the limits
of distributions of $\langle s_j\rangle_N$, which, after some
minor technical efforts, lead to correctly defined random
infinite-volume Gibbs states. The same problem exists in the
spherical model, and, most likely, in such often considered models
as the Ising model and $O(n)$ models. Namely, for $b=0$,
$\lim_{n\to\infty}\langle x_j\rangle_n$ does not exist almost
surely, although it exists in distribution. The results of the
present paper show that switching on a homogeneous boundary field
rectifies the problem with almost sure convergence. Namely, for
$b\neq0$, $\lim_{n\to\infty}\langle x_j\rangle_n$ exist almost
surely, which (together with convergence of higher correlation
functions) means that the corresponding limit Gibbs state exists
for almost all realizations of the random field $\{\mbox{\boldmath
$h$}_j,\,j\in Z^d\}$.

At approximately the same time, Newman and Stein \cite{ns92}
pointed out that the absence of convergence of local thermodynamic
averages, like $\langle s_j\rangle_N$, is a natural occurrence in
many disordered systems. They call this phenomenon the chaotic
size dependence. Somewhat later, Newman and  Stein also proposed
their own solution of the problem with infinite-volume Gibbs
states. Instead of looking at distributions of local averages like
$\langle s_j\rangle_N$ they choose to look at the empirical
distributions
\[
{\cal F}_N(y)\equiv N^{-1}\#\{k\in\{1,2,\ldots,N\}:\langle
s_j\rangle_k\leq y\},
\]
for a fixed realization of randomness. Assuming ergodicity we have
\[
\lim_{N\to\infty}{\cal F}_N(y)=\lim_{N\to\infty}\Pr\left[\langle
s_j\rangle_N\leq y\right],
\]
hence, both constructions provide the same result: a random
infinite-volume Gibbs state. Newman and Stein call the random
Gibbs state the metastate.

The authors of the paper \cite{ah96} investigated self-averaging
using the ideas of renormalization group theory. They concluded
that there are universality classes of models within which a
particular non-self-averaging thermodynamic observable has the
same distribution in the thermodynamic limit. The results of the
present paper indicate that the conclusion of the paper
\cite{ah96} looks plausible, at least for the magnetization.
Indeed, according to Eq.\ (\ref{mgn}) the magnetization of the
spherical model obtains the values $\pm m^*$ with probability
$\frac{1}{2}$, where $m^*$ is the spontaneous magnetization. The
magnetization of the Curie-Weiss model and, most likely, of
disordered finite-dimensional Ising models has the same
distribution, see \cite{apz92}. One can also guess that the
magnetization of various disordered $O(n)$ models is uniformly
distributed over an $n$-dimensional sphere. On the other hand, we
also saw that the distribution of the magnetization is
highly-sensitive to {\em symmetry-breaking} perturbations. Indeed
an arbitrarily weak symmetry-breaking boundary field restores
self-averaging, that is, changes a non-degenerate distribution to
a degenerate one. Although that fact rather goes along with than
contrary to the lines of renormalization group argument.

The susceptibility of the spherical model
\[
\chi_n=\beta n^d\left[\left\langle\left(\frac{1}{n^d} \sum_{j\in
V_n}x_j\right)^2 \right\rangle-\left(\left\langle\frac{1}{n^d}
\sum_{j\in V_n}x_j \right\rangle\right)^2\right]\equiv \beta
n^d\,\mbox{t-Var}(m_n),
\]
can be easily found from Eqs.\ (\ref{mgnb}) and (\ref{mgn}). If
$b\neq0$, then (when properly normalized) the susceptibility is
self-averaging
\[
\chi_n\sim\frac{n^2}{4 J\zeta_0}\left(1-\frac{\tanh\sqrt{\zeta_0}}
{\sqrt{\zeta_0}}\right),
\]
while if $b=0$ then the susceptibility is not a self-averaging
observable
\begin{equation}
\chi_n\sim
\frac{n^{d/2}}{|\varphi_{(1,1,\ldots,1)}|}\frac{8}{\pi^2}\sqrt{\left(1-
\frac{\beta_c}{\beta}\right)\frac{W_d^{(1)}(d)}{2\beta_c J}}.
\label{susc}
\end{equation}

The susceptibility of various 3D disordered models was studied
intensively using Monte-Carlo simulations since mid-90s, see,
e.g.\ \cite{r95,wd95}. The histograms obtained in \cite{r95,wd95}
suggest that the distribution of the susceptibility is not normal,
positively skewed, and has heavy tails. The distribution of the
susceptibility given by Eq.\ (\ref{susc}) has the same properties,
and thus, to some extend, explains the results of Monte-Carlo
simulations. It has been suggested in the paper \cite{ah96} that
the distribution of susceptibility should be the same within
universality classes. Since Eq.\ (\ref{susc}) is the asymptotics
of Eq.\ (\ref{vart}) at the pole $z=\lambda_{\rm max}$ it is not
unreasonable to expect the universality of the distribution of
$\chi_n$ for a certain class of models. Although it is tempting to
speculate that $O(n)$ models might belong to the universality
class, nevertheless, the results of the present paper do not
indicate neither how wide the universality class is, nor which
models possibly belong to this class.

In conclusion, various disordered models have been intensively
studied recently either numerically or using various heuristic
approaches like, for instance, the renormalization group. The
present paper derives explicitly distributions of various
thermodynamic quantities within a non-trivial disordered
finite-dimensional model --- the spherical model in a random
field. The author hopes that the paper is helpful for
understanding the conclusions of heuristic theories, and for
interpreting the results of Monte-Carlo simulations.

\end{document}